\documentclass[a4paper,12pt]{article}
\usepackage{amsfonts,slashed}
\usepackage{url}
\usepackage{latexsym}
\usepackage{amsfonts}
\usepackage{epsfig}
\usepackage{latexsym,amssymb}
  \usepackage{amsmath,amssymb,amsthm}
\usepackage{ifpdf}
\ifx\pdfoutput\undefined
   \pdffalse
   \usepackage{cite}
 \else
   \pdfoutput=1
   \pdftrue
  \usepackage[pdftex]{hyperref}
\fi

\numberwithin{equation}{section}
\def\be{\begin{equation}}
\def\ee{\end{equation}}
\def\ba{\begin{array}}
\def\ea{\end{array}}

\newcommand{\bea}{\begin{eqnarray}}
\newcommand{\eea}{\end{eqnarray}}

\def\ii{{\rm i}}

\newcommand{\bbox}{\lower.2ex\hbox{$\Box$}}

\def\bfone{\relax{\rm 1\kern-.35em 1}}
\def\bfzero{\relax{\rm 0\kern -.45 em 0}}

\usepackage{tikz}
\usetikzlibrary{trees}
\usetikzlibrary{positioning,shadows,arrows,decorations.pathmorphing,decorations.markings}
\tikzstyle{block}=[draw opacity=0.7,line width=1.4cm]

\begin{document}
\begin{flushright}
\vskip 18mm
\end{flushright}
\begin{center}
{\bf\LARGE Hidden Gauge Structure of Supersymmetric Free Differential Algebras} \\
\vskip 2 cm
{\bf \large Laura Andrianopoli$^{1,2}$,  Riccardo D'Auria$^{1}$
 and Lucrezia Ravera$^{1,2}$}
\vskip 8mm
 \end{center}
\noindent {\small $^1$ DISAT, Politecnico di Torino, Corso Duca
    degli Abruzzi 24, I-10129 Turin\\
    $^{2}$  \it Istituto Nazionale di
    Fisica Nucleare (INFN) Sezione di Torino, Italy
}

\vskip 1 cm

\begin{center}
\small {\bf Abstract}
\end{center}
 The aim of this paper is to clarify the role of the nilpotent fermionic generator $Q'$ introduced in \cite{D'Auria:1982nx} and appearing in the hidden supergroup underlying the free differential algebra (FDA) of D=11 supergravity.

 We give a  physical explanation of its  role by looking at the gauge properties of the theory.
 We find that its presence
 is necessary, in order that  the extra 1-forms of the hidden supergroup give rise to the correct gauge transformations of the $p$-forms of the FDA.
 This interpretation is actually valid for any supergravity containing antisymmetric tensor fields, and
 any supersymmetric  FDA  can always be traded for a hidden Lie superalgebra containing extra fermionic nilpotent generators.

 As an interesting example we   construct   the  hidden superalgebra associated with the FDA of $\mathcal{N}=2$, $D=7 $ supergravity. In this case we are able to parametrize the mutually non local $2$- and $3$-form $B^{(2)}$ and $B^{(3)}$ in terms of hidden $1$-forms and find that supersymmetry and gauge invariance require  in general the presence of
 two nilpotent fermionic generators in the hidden algebra.

We propose that our approach, where all the invariances of the FDA are expressed as Lie derivatives of the $p$-forms in the hidden supergroup manifold, could be an appropriate framework to discuss theories defined in enlarged versions of superspace recently considered in the literature, such us double field theory and its generalizations.


\vskip 1 cm
\vfill
\noindent {\small{\it
    E-mail:  \\
{\tt laura.andrianopoli@polito.it}; \\
{\tt riccardo.dauria@polito.it}; \\
{\tt lucrezia.ravera@polito.it}}}
   \eject
   \numberwithin{equation}{section}
\section{Introduction}

Supergravity theories
 in various space-time dimensions $4\leq D\leq 11$ have a bosonic field content that generically includes, besides the metric and a set of 1-form gauge potentials, also  $(p+1)$-form gauge potentials of various $p\leq 9$, and they are therefore appropriately discussed in the context of Free Differential Algebras (FDA in the following). This is also required from superstring theories, where the higher form potentials are related to the NS-NS and R-R sectors of the different superstring theories \cite{string}.

 Early after the discovery of  Supergravity,  the action of $D=11$ supergravity was first constructed in \cite{Cremmer:1978km}. It has a bosonic field content given by the metric $g_{\mu\nu}$ and a 3-index  antisymmetric tensor $A_{\mu\nu\rho}$ ($\mu,\nu,\rho, \cdots =0,1,\cdots,D-1$), together with a single Majorana gravitino $\Psi_\mu$ in the fermionic sector.

  The structure of this same theory was reconsidered in \cite{D'Auria:1982nx} in the framework of   FDAs using the superspace geometric approach. In this setting, its bosonic sector includes, besides the supervielbein $\{V^a,\Psi\}$, a 3-form potential $A^{(3)}$, with field-strength $F^{(4)}= dA^{(3)}$ (modulo gravitino 1-form  bilinears), together with its Hodge-dual $F^{(7)}$, defined such that its space-time components are related to the ones of the  4-form by $F_{\mu_1 \cdots \mu_7}= \frac 1{84} \epsilon_{\mu_1\cdots \mu_7\nu_1\cdots \nu_4} F^{\nu_1\cdots \nu_4}$; this amounts to say that it is associated with a 6-form potential $B^{(6)}$ in superspace. The on-shell closure of the supersymmetric theory relies on 3-fermions Fierz identities and requires $F^{(7)}= dB^{(6)}-15 A^{(3)}\wedge F^{(4)}$ (modulo fermionic currents).

   In the same paper the supersymmetric FDA was also investigated in order to see whether the FDA formulation  could be interpreted in terms of an ordinary Lie superalgebra in its dual Maurer-Cartan formulation.
   Actually, this was proven to be true and the existence of a  superalgebra underlying the theory was presented for the first time.

   This superalgebra includes as a subalgebra the super-Poincar\'e algebra of the eleven dimensional theory, but it also contains two extra bosonic generators $Z^{ab}, Z^{a_1 \cdots a_5}$ ($a,b,\cdots = 0,1,\cdots 10$), which commute with the 4-momentum $P_a$, while  having appropriate commutators with the eleven dimensional Lorentz generators $J_{ab}$. \footnote{They are indeed 1-forms valued in the antisymmetric tensor represemtations of $\rm SO(1,10)$.  } In the following,  generators that commute with all the superalgebra but the Lorentz generators will be named ``\emph{almost central}".
   Furthermore, to close the algebra, an extra nilpotent fermionic generator called $Q'$ must be included.

 Besides the standard Poincar\'{e} Lie algebra, the superalgebra associated with $D=11$ supergravity has the following structure of (anti)commutators:
   \begin{eqnarray}
     \{Q,Q\}&=& -\ii C\Gamma^a P_a - \frac{ 1}{2} C\Gamma_{ab} Z^{ab} - \frac{ \ii}{5!} C\Gamma_{a_1 \cdots a_5} Z^{a_1 \cdots a_5}\,;\label{democracy}\\
    \left[Q,P_a \right]&\propto&   \Gamma_a Q' \label{newdemo1}\,;\\
   \left[Q,Z^{ab}\right]&\propto &   \Gamma^{ab} Q' \label{newdemo2}\,;\\
   \left[Q,Z^{a_1 \cdots a_5}\right]&\propto &   \Gamma^{a_1 \cdots a_5} Q' \label{newdemo3}\,;\\
   \{Q',Q'\}&=&0\,;\label{nilpot}
   \end{eqnarray}
   together with
   \begin{equation}
    \hskip -3mm \left[J_{ab},Z^{cd}\right]\,\propto\, \delta_{[a}^{[c}\ \eta_{{b]}l}\ Z^{{d]} l}\,;\quad \left[J_{ab},Z^{c_1\cdots c_5}\right]\,\propto \,\delta_{[a}^{[c_1} \ \eta_{{b]}l}\  Z^{  c_2\cdots c_5] l}\,;\quad \left[J_{ab},Q'\right]\,\propto\, \Gamma_{ab}Q',
   \end{equation}
   the other (anti)commutation relations being zero. The precise relations  are reported in Section \ref{11D}. Here and in the following we shall refer to a superalgebra descending from a given FDA as a \emph{hidden superalgebra}. Note that the set of generators $\{Z^{ab},Z^{a_1\cdots a_5}, Q'\}$,  extending the super-Poincar\'e Lie algebra to the
    hidden superalgebra written above, actually span
  an abelian ideal of it. They will also be referred to as  \emph{hidden generators}.

   Let us remark that the anticommutation relation (\ref{democracy}) generalizes to almost  central charges  the central extension of the supersymmetry algebra \cite{Haag:1974qh}, which was shown in \cite{Witten:1978mh} to be associated with  topologically non trivial configurations of the bosonic fields.  The possible extension (\ref{democracy}) of the supersymmetry algebra,  for supergravity theories in $D>4$ dimensions, was later widely considered (see in particular \cite{vanHolten:1982mx}-\cite{Abraham:1990nz}). After the discovery of D$p$-branes as sources for the R-R gauge potentials \cite{Polchinski:1995mt} and the ensuing understanding of the duality relation between eleven dimensional supergravity and Type IIA theory in ten dimensions, the bosonic generators $Z^{ab}, Z^{a_1 \cdots a_5}$ were understood as $p$-brane charges, sources of the dual potentials $A_{(3)}$ and $B_{(6)}$ \cite{Hull:1994ys,Townsend:1995gp}, and eq. (\ref{democracy}) was then interpreted as the natural generalization of the supersymmetry algebra in higher dimensions, in the presence of non-trivial topological extended sources (black $p$-branes).
 \vskip 4mm
 However,  the structure of the full superalgebra, given in eq.s (\ref{democracy}) - (\ref{nilpot}), which is hidden in the superymmetric D=11 FDA,
 besides the almost central charges $Z^{ab}$ and $Z^{a_1\cdots a_5}$, also requires for its consistency (closure of the super-Jacobi identities) the presence of an extra fermionic nilpotent charge, $Q'$, as shown in reference \cite{D'Auria:1982nx}.
This fact  is not a peculiarity of the eleven dimensional theory, but is fully general, and, as we will extensively discuss in this paper, a hidden superalgebra underlying the supersymmetric FDA \emph{containing at least one nilpotent fermionic generator} can be constructed for each supergravity theory where antisymmetric tensor fields are present.

The role played by the extra fermionic generator $Q'$ and
its group-theoretical and physical meaning,  corresponding to the  non-trivial contributions  (\ref{newdemo1}) - (\ref{nilpot}), was much less investigated with respect to that of the almost central charges. The most relevant contributions that we are aware of were given first in \cite{vanHolten:1982mx} and then
in particular in \cite{Bandos:2004xw}, where the results in \cite{D'Auria:1982nx} were further analyzed and generalized. However, the physical meaning of $Q'$ remained obscure, at our knowledge.

 Actually, the consistency of the $D=11$ theory, that is the closure of the supersymmetric FDA and of its hidden superalgebra, fully relies on 3-fermion Fierz identities obeyed by the gravitino 1-forms, and it crucially requires the presence of the nilpotent spin-3/2 field $\eta$ associated with the fermionic charge $Q'$.
   Three-gravitini Fierz-identities are  at the heart of the closure of all lower dimensional supergravities, and in particular of those  based on FDA's. As a consequence of this, almost central-extended hidden superalgebras,  including extra nilpotent fermionic generators as necessary ingredients, should underly all the supergravity theories based on FDAs, as we have explicitly checked in various supergravity models with $6\leq D\leq 9$.

It is the aim of the present paper to further investigate the superalgebra hidden in all the supersymmetric FDAs
 and to clarify the role played by its the bosonic and fermionic  generators.
 In particular, we will analyze  in detail the gauge structure of the supersymmetric FDA in eleven dimensions in relation to its hidden gauge superalgebra, and then we will consider a specific case in lower dimensions (we will choose minimal supergravity in D=7) to test the universality of the construction and to investigate possible extensions of the underlying superalgebra of \cite{D'Auria:1982nx}.

    The main result of our paper is to disclose the physical interpretation of the fermionic hidden generator $Q'$. We will show that  it has a topological meaning, since it controls the gauge structure of the FDA once it is expressed in terms of 1-forms. We will also find that in general more than one nilpotent fermionic generator are necessary to construct  the fully extended superalgebra hidden in the supersymmetric FDA. This will be the case in particular of the minimal supersymmetric $D=7$ FDA, which we will analyze in some detail.

Considering now the bosonic hidden generators of the hidden algebra  (we will call $H_b$ the corresponding  tangent space directions of the hidden group manifold), we will show that they are associated with internal diffeomorphisms of the supersymmetric FDA in $D$ dimensions. More precisely, once a $p$-form $A^{(p)}$ of the FDA is parametrized in terms of the hidden 1-forms, contraction of $A^{(p)}$ along a generic tangent vector $\vec z\in H_b$ gives a $(p-1)$-form gauge parameter, and  the Lie derivative of the FDA along a tangent vector $\vec z$ gives a gauge transformation leaving the FDA invariant.

This construction is not limited to the eleven dimensional FDA. In particular, it is  interesting to
 consider ten dimensional Type IIA supergravity, which  naturally descends from  the $D=11$ theory. Its FDA includes the  2-form NS-NS field $B^{(2)}$, also appearing in all superstring-related supergravities, which has a natural understanding in terms of the antisymmetric 3-form $A^{(3)}$ of {D=11} supergravity.  The corresponding hidden 1-form field, $B_a$, has an associated charge $Z^a$  which carries a Lorentz-index, contravariant with respect to the one carried by the translation generator $P_a$. It follows that in the fully extended hidden superalgebra in any $D\leq 10$, $P_a$ and $Z^a$ appear on the same footing and the action of the hidden superalgebra in this case includes automorphisms interchanging them. When some of the space-time directions are compactified on circles, these automorphisms are naturally associated with T-duality transformations interchanging  momentum with winding in the compact directions.

As we are going to discuss in the following, the structure outlined above is strongly reminiscent of the one described in the framework of generalized geometry \cite{Hitchin:2004ut}-\cite{Grana:2008yw} and its extensions to M-theory \cite{Hull:2007zu}-\cite{Coimbra:2012af}, double field theory \cite{Hull:2009mi}-\cite{Hull:2014mxa}  and exceptional field theory \cite{Hohm:2013pua}-\cite{Hohm:2014qga}.
We expect that our formalism could be  useful in this context.

To clarify the crucial role played by the nilpotent hidden fermionic generators  for the consistency of the hidden superalgebra, we will consider a singular limit where  the associated spinor 1-form $\eta$ satisfies $\eta\to 0$. In this limit the supersymmetric FDA parametrized in terms of 1-forms becomes ill defined: indeed the exterior forms $A^{(p)}$ are gauge fields, since they include ``longitudinal" unphysical  directions corresponding to the gauge freedom $A^{(p)} \to A^{(p)} + d\Lambda^{(p-1)}$. In the limit $\eta \to 0$,  the unphysical degrees of freedom $\Lambda^{(p-1)}$ get mixed with the physical directions of the superspace, and all the generators of the hidden superalgebra act as generators of external diffeomorphisms.
On the contrary, when $\eta\neq 0$  the hidden supergroup acquires a principal fiber bundle structure  and allows to separate, in a dynamical way, the physical directions of superspace, generated by the super-vielbein $(V^a,\Psi)$, from the other directions, belonging to the fiber of superspace, in such a way as to recover the gauge invariance of the FDA.

In this paper we will limit ourselves to consider  the FDA, and its underlying supergroup corresponding to the ground state of the supergravity theory, also referred to as the ``vacuum", which is  defined by the condition that all the supercurvatures vanish, so that only the topological structure and  the symmetries of the theory emerge. As is usual in supersymmetric theories, they include, besides the local symmetries which can be realized at the lagrangian level, also non-perturbative symmetries, associated with mutually non-local generators.
We will not consider here the full dynamical content of the theory out of the vacuum, where the simultaneous presence of mutually non-local electric and magnetic $p$-forms is forbidden at the lagrangian level\footnote{ Some progress in this topic has been obtained in reference \cite{Bandos:2004xw}}. For the D=7 theory under consideration, we will show that it is however possible to find two inequivalent ``Lagrangian subalgebras"   of the hidden superalgabra, which only include mutually local fields and which should be relevant for the Lagrangian description of the interacting theory.
Actually, each of them includes, as hidden fermionic generators, only one of the two nilpotent spinors.

 The paper is organized as follows:

In Section \ref{11D} we will review, in a critical way, the various steps of the construction of the superalgebra hidden in eleven dimensional supergravity, following \cite{D'Auria:1982nx}.

 Then, in Section \ref{CE} we will analyze in detail the gauge structure of the hidden superalgebra, discussing in particular the role of the nilpotent generator in the D=11 supersymmetric FDA.

 In Section \ref{7D} we will  focus our study on the minimal $D=7,\,\,\mathcal{N}=2$ supergravity theory, whose FDA is particularly rich since it includes, besides a triplet of gauge vectors $A^x$, a 2-form $B^{(2)}$, a 3-form $B^{(3)}$ related to $B^{(2)}$ by Hodge-duality of the corresponding field strengths, and a triplet of 4-forms $A^{x|(4)}$ related to $A^{x}$ by Hodge-duality of the corresponding field strengths. This theory can be obtained by dimensional reduction, on a four-dimensional compact manifold, preserving only half of the supersymmetries,  from $D=11$ supergravity.
 We will provide the parametrization in terms of 1-forms of the mutually non local fields $B^{(2)}$ and $B^{(3)}$,  finding the corresponding superalgebra hidden in the supersymmetric FDA. Actually in this case we will find that \emph{two extra nilpotent fermionic generators are required} for the closure of the fully extended hidden superalgebra.

In Section \ref{11d7d} we will consider the dimensional reduction of the $D=11$ FDA to $D=7$ on an orbifold $T^4/Z_2$, showing the conditions under which the seven dimensional model studied in Section \ref{7D} could be obtained by dimensional reduction of the eleven dimensional model of Section \ref{11D}.

The main body of the  paper ends  in Section \ref{concl} with some concluding remarks. Our notations and conventions, together with some technical details, can be found in the Appendices.

\section{Review of the eleven dimensional hidden superalgebra}\label{11D}
As said in the introduction, the $D=11$ theory, first constructed in \cite{Cremmer:1978km}, was reformulated in ref. \cite{D'Auria:1982nx} using a geometric superspace approach, in terms of a supersymmetric FDA. \footnote{In the original paper \cite{D'Auria:1982nx} the FDA was referred to as Cartan Integrable System (CIS), since the authors were unaware of the previous work by Sullivan \cite{Sullivan} who actually introduced the mathematical concept of FDA  to  which the CIS are equivalent. }
In this context the bosonic vielbein $V^a$ ($a=0,1,\cdots , 10$), together with the gravitino 1-form $\Psi$, span a basis of the cotangent superspace $K\equiv\{V^a,\Psi\}$, where also the superspace 3-form $A^{(3)}$, whose pull-back on space-time is $A_{\mu\nu\rho}$, is defined.

Actually, it was stressed there that besides  the simplest FDA including as exterior form only $A^{(3)}$,  one can fully  extend the FDA to include also a (\emph{magnetic}) 6-form potential $B^{(6)}$, related   to $A^{(3)}$ by  Hodge-duality of the corresponding field-strengths. More precisely, the supersymmetric FDA, which defines the ground state of the theory, is given by the vanishing of  the following set of supercurvatures:
\begin{eqnarray}
R^{ab}&\equiv& d\omega^{ab} - \frac 12 \omega^{ac}\wedge \omega^{bd}\eta_{cd}=0\,,\label{FDA11omega}\\
T^a&\equiv& D V^a - \frac{\ii}{2}\overline{\Psi}\wedge \Gamma^a \Psi =0\,,\label{FDA11v} \\
\rho&\equiv&D \Psi=0\,,\label{FDA11psi}\\
F^{(4)} &\equiv&dA^{(3)} - \frac{1}{2}\overline{\Psi}\wedge \Gamma_{ab}\Psi \wedge V^a \wedge V^b =0\,,\label{FDA11a3} \\
F^{(7)}&\equiv&dB^{(6)} - 15 A^{(3)}\wedge  dA^{(3)} -\frac{\ii}{2}\overline{\Psi}\wedge \Gamma_{a_1\cdots a_5}\Psi \wedge V^{a_1} \wedge \cdots V^{a_5}=0\,,\label{FDA11b6}
\end{eqnarray}
where $D$ denotes the eleven dimensional Lorentz-covariant derivative and its closure $d^2=0$ is a consequence of 3-fermions Fierz identities in eleven dimensions (see Appendix \ref{fierz}). \footnote{In the ground state the spin-1/2 fields are zero by Lorentz invariance and the scalar fields are constant (they can be set to zero).}
The  interacting theory (out of the ground state),
 including the field equations, is obtained in this setting through a straightforward procedure \cite{D'Auria:1982nx},\cite{Castellani:1991et}, corresponding to  introducing a non-vanishing value to the super-curvatures defined in the left-hand side of the FDA, and  given respectively by the super Riemann 2-form $R^{ab}$, the supertorsion $T^a$, the gravitino super field-strength $\rho$,  the 4-form $F^{(4)}$ and its Hodge-dual $F^{(7)}$.
We will not further elaborate on this, here, since the topological structure of the theory, which will be the object of the present investigation, is fully catched by the ground state FDA.

 The authors of \cite{D'Auria:1982nx} asked themselves whether one could trade the FDA structure on which the theory is based with an ordinary   Lie superalgebra, written in its dual Cartan form, that is in terms of 1-form gauge fields which turn out to be valued in non trivial tensor representations of  Lorentz group $\rm{SO}(1,10)$.
 This would allow to disclose the fully extended superalgebra hidden in the supersymmetric FDA.

 It was found that this is indeed possible by associating, to the   forms $A^{(3)}$ and $B^{(6)}$, the bosonic 1-forms $B_{ab}$ and $B_{a_1 \cdots a_5}$,  in the antisymmetric representations of $\rm{SO}(1,10)$, whose Maurer-Cartan equations are:
 \begin{eqnarray}
 D B_{a_1a_2} & = & \frac{1}{2}\overline{\Psi}\wedge \Gamma_{a_1a_2}\Psi , \\
D B_{a_1...a_5}& = & \frac{\ii}{2} \overline{\Psi}\wedge \Gamma_{a_1...a5}\Psi\,,\label{also}
\end{eqnarray}
$D$ being the  Lorentz-covariant derivatives.
  In particular, they presented a general decomposition of the 3-form $A^{(3)}$ in terms of the  1-forms $B_{ab}$ and  $B_{a_1...a_5}$, by requiring the Bianchi identities in superspace of the 3-form, $d^2 A^{(3)}=0$, to be satisfied also when $A^{(3)}$ is decomposed in terms of the 1-forms $B_{ab}$ and $B_{a_1...a_5}$.
   Actually, it was shown that this program can be accomplished if and only if, together with  the newly introduced bosonic 1-form fields, one also introduces an extra \emph{spinor} 1-form $\eta$, satisfying:
 \begin{eqnarray}
 D \eta & = & \ii E_1 \Gamma_a \Psi \wedge V^a + E_2 \Gamma^{ab}\Psi \wedge B_{ab}+ \ii E_3 \Gamma^{a_1...a_5}\Psi \wedge B_{a_1...a_5}\,.\label{Deta}
\end{eqnarray}

 They found that the most general solution enjoying the above requirements has the following form \footnote{Here, and in the following, with $B_{a_1...a_{p-1}}^{\;\;\;\;\;\;\;\;\;\;\;\;\;\;b}$ we generally mean $B_{a_1...a_p}\eta^{b a_p}$, where $\eta_{ab}=(+,-,\cdots,-)$ denotes the Minkowski metric}:
\begin{eqnarray}\label{29}
A^{(3)} & = & T_0 B_{ab} \wedge V^a \wedge V^b + T_1 B_{a b}\wedge B^{b} _{\;c}\wedge B^{c a}+ \nonumber\\
& +& T_2 B_{b_1 a_1...a_4}\wedge B^{b_1}_{\; b_2}\wedge B^{b_2 a_1...a_4}+ T_3 \epsilon_{a_1...a_5 b_1...b_5 m}B^{a_1...a_5}\wedge B^{b_1...b_5}\wedge V^m + \nonumber\\
& +& T_4 \epsilon_{m_1...m_6 n_1...n_5}B^{m_1m_2m_3p_1p_2}\wedge B^{m_4m_5m_6p_1p_2}\wedge B^{n_1...n_5} + \nonumber\\
& + & \ii S_1 \overline{\Psi}\wedge \Gamma_a \eta \wedge V^a + S_2 \overline{\Psi}\wedge \Gamma^{ab}\eta \wedge B_{ab}+ \ii S_3 \overline{\Psi}\wedge \Gamma^{a_1...a_5}\eta \wedge B_{a_1...a_5}\,,\label{a3par}
\end{eqnarray}
where the requirement that $A^{(3)}$ in (\ref{a3par}) satisfies eq. (\ref{FDA11a3})
fixes the free constants $T_i$, $S_j$ in terms of the structure constants $E_1,E_2,E_3$.
Actually, the consistence of the theory also requires the $d^2$ closure of the newly introduced fields  $B_{ab}$, $B_{a_1\cdots a_5}$ and $\eta$. For the two bosonic 1-form fields the  $d^2$ closure is obvious in the ground state, because of the vanishing of the curvatures $R^{ab}$ and $\rho$, while on $\eta$ it requires the further condition:
\begin{equation}\label{integrability11}
E_1+10E_2-720E_3=0\,.
\end{equation}
The final result is:
\begin{eqnarray}
T_0 &=& \frac{120 {E_3}^2}{({E_2}-60{E_3})^2}+\frac{1}{6} , \;\;\; T_1 \; = \; -\frac{{E_2} ({E_2}-120 {E_3})}{90 ({E_2}-60 {E_3})^2} , \;\;\; T_2 \;=\; -\frac{5 {E_3}^2}{({E_2}-60 {E_3})^2},  \nonumber\\
T_3 &=& \frac{{E_3}^2}{120 ({E_2}-60 {E_3})^2}, \;\;\; T_4 \;=\; -\frac{{E_3}^2}{216 ({E_2}-60 {E_3})^2} , \;\;\; S_1 \;=\; \frac{{E_2}-48 {E_3}}{24({E_2}-60 {E_3})^2}, \nonumber\\
S_2 &=&-\frac{{E_2}-120 {E_3}}{240 ({E_2}-60 {E_3})^2}, \;\;\; S_3\;=\; \frac{{E_3}}{240 ({E_2}-60{E_3})^2}, \nonumber\\
 E_1 &=& -10 ({E_2}-72{E_3}) .\label{11dsol}
\end{eqnarray}
 where the constants $E_1,E_2,E_3$ define new structure constants of the hidden super-algebra.

In \cite{D'Auria:1982nx} the first coefficient $T_0$ was arbitrarily fixed to $T_0=1$ giving only 2 possible solutions for the set of parameters $\{T_i,S_j,E_k\}$. It was pointed out later in \cite{Bandos:2004xw} that this restriction can be relaxed thus giving the general solution (\ref{11dsol}). Indeed, as  observed in the quoted reference, one of the $E_i$ can be reabsorbed in the normalization of $\eta$, so that, owing to the relation (\ref{Deta}), we are left with one free parameter, say $E_3/E_2$.\footnote{In reference \cite{Bandos:2004xw} their free parameter $s$ is different from ours and is related to $E_3/E_2$=$\rho$ by the relation $\frac{120\rho -1}{90\left(60\rho-1\right)^2} =\frac{2(3+s)}{15s^2}$\,.} The details of the calculation are reported in Appendix \ref{coeff11D}, where also some misprints of \cite{D'Auria:1982nx}, in part recognized already in \cite{Bandos:2004xw}, are corrected.

The full Maurer-Cartan equations of the hidden algebra (in dual form) are then:
 \begin{eqnarray}
 d\omega^{ab}&=&\frac 12 \omega^{ac}\wedge \omega^{bd}\eta_{cd}\\
 D V^a &=& \frac{\ii}{2}\overline{\Psi}\wedge \Gamma^a \Psi, \\
D \Psi&=&0, \\
 D B_{a_1a_2} & = & \frac{1}{2}\overline{\Psi}\wedge \Gamma_{a_1a_2}\Psi , \\
D B_{a_1...a_5}& = & \frac{\ii}{2} \overline{\Psi}\wedge \Gamma_{a_1...a_5}\Psi,\\
 D \eta & = & \ii E_1 \Gamma_a \Psi \wedge V^a + E_2 \Gamma^{ab}\Psi \wedge B_{ab}+ \ii E_3 \Gamma^{a_1...a_5}\Psi \wedge B_{a_1...a_5}\,. \label{deta}
\end{eqnarray}
%

Let us finally write down the hidden superalgebra in terms of generators closing a set of (anti)com\-mu\-ta\-tion relations.
 For a generic set of 1-forms $\sigma^\Lambda$ satisfying the Maurer-Cartan equations:
$$d\sigma^\Lambda = -\frac 12 C^\Lambda_{\ \ \Sigma\Gamma}\sigma^\Sigma \wedge\sigma^\Gamma \,,$$
in terms of structure constants $C^\Lambda_{\ \ \Sigma\Gamma}$, this is performed by introducing a set of dual generators $T_\Lambda$ satisfying
\begin{eqnarray}
\sigma^\Lambda(T_\Sigma) = \delta^\Lambda_\Sigma\,;\qquad
d\sigma^\Lambda (T_\Sigma,T_\Gamma)=  C^\Lambda_{\ \ \Sigma\Gamma}
\end{eqnarray}
so that the $\{T_\Lambda\}$ close the algebra $[T_\Sigma, T_\Gamma]=  C^\Lambda_{\ \ \Sigma\Gamma} T_\Lambda$.

In the case at hand, the 1-forms $\sigma^\Lambda$ are
\begin{equation}
\sigma^\Lambda\equiv\{V^a, \Psi, \omega^{ab}, B_{ab}, B_{a_1...a_5}, \eta\}\,.
 \label{sigma11d}
\end{equation}
To recover the superalgebra  in terms of (anti)-commutators of the dual Lie superalgebra generators:
\begin{equation}T_\Lambda \equiv\{P_a, Q, J_{ab},  Z^{ab}, Z^{a_1...a_5}, Q'\}\,,\label{t11d}
\end{equation}
we use the duality between 1-forms and generators defined by the usual conditions:
\begin{eqnarray}
&V^a(P_b)= \delta^a_b\,,\quad \Psi(Q) =\bfone\,,\quad \omega^{ab}(J_{cd})=  {2}\delta^{ab}_{cd}\,,&\nonumber\\
& B^{ab}(Z_{cd})= {2}\delta^{ab}_{cd}\,,\quad B^{a_1...a_5}(Z_{b_1...b_5})=  {5!}\delta^{a_1...a_5}_{b_1...b_5}\,,\quad \eta(Q') =\bfone &
\end{eqnarray}
where $\bfone$ denotes unity in the spinor representation.
The $D=11$ FDA then corresponds to the following hidden contributions to the superalgebra (besides the Poincar\'e algebra):
\begin{eqnarray}
\lbrace Q,\bar Q \rbrace &=& -\left(\ii \Gamma^a P_a + \frac 12 \Gamma^{ab}Z_{ab}+ \frac {\ii}{5!} \Gamma^{a_1...a_5}Z_{a_1...a_5}\right)\,, \label{qq11}\\ \nonumber
\lbrace Q',\bar Q' \rbrace &=& 0\,,\\ \nonumber
[Q, P_a] &=& -2 \ii E_1 \Gamma_a Q'\,,\\ \nonumber
[Q, Z^{ab}] &=&-4 E_2 \Gamma^{ab}Q' \,, \\ \nonumber
[Q, Z^{a_1...a_5}] &=&- 2 \,(5!) \ii E_3 \Gamma^{a_1...a_5}Q'\,, \\ \nonumber
[J_{ab}, Z^{cd}]&=&-8 \delta^{[c}_{[a}Z_{b]}^{\ d]}\,,\\ \nonumber
[J_{ab}, Z^{c_1\dots c_5}]&=&- 20 \delta^{[c_1}_{[a}Z^{c_2\dots c_5]}_{b]}\,,\\ \nonumber
[J_{ab}, Q]&=&- \Gamma_{ab} Q\,,\\ \nonumber
[J_{ab}, Q']&=&- \Gamma_{ab} Q'\,.
\end{eqnarray}
All the other commutators (beyond the Poincar\'e part) vanishing.
As said before, the $E_i$ satisfy equation (\ref{integrability11}) and one of them can be reabsorbed in the normalization of the $\eta$ $1$-form. \footnote{The  closure of the superalgebra under (super)- Jacobi identities
is a consequence of the $d^2$-closure of the Maurer-Cartan 1-forms equations.}

Finally, let us recall that the presence of the bosonic hidden 1-forms $B_{ab}, B_{a_1...a_5}$ in the relation (\ref{qq11}), which generalizes the centrally extended supersymmetry algebra of \cite{Witten:1978mh} (where the central generators were  associated with electric and magnetic charges), has in fact a topological meaning.  This  was recognized in \cite{Achucarro:1987nc} and \cite{deAzcarraga:1989mza},   where it was shown they to be associated with extended objects  (2-brane and 5-brane) in space-time. In particular in reference  \cite{deAzcarraga:1989mza} it was shown that quite generally such p-forms must be present in any dimensions, their associated (almost) central charges  appearing   in  the supersymmetry algebra.  As we shall see, this in fact occurs in the minimal $D=7$ theory that will we shall analyze in section \ref{7D}. The results of \cite{D'Auria:1982nx}, and those of \cite{Achucarro:1987nc} and \cite {deAzcarraga:1989mza} can thus be considered an important extension of the property found  in \cite{Witten:1978mh}.

On the other hand, the fact that the  supersymmetry algebra, once extended to its hidden superalgebra, requires the presence of extra spinor generators, was not discussed in \cite{Achucarro:1987nc,deAzcarraga:1989mza}. As we are going to discuss in the next Section, the presence of nilpotent fermionic charges in the hidden sector has instead a crucial role for the consistence of the FDA in superspace.

\section{FDA Gauge Structure and Supergravity}\label{CE}
The aim of this section is to analyze in detail the hidden gauge structure of the FDA of  D=11 supergravity, when the exterior $p$-forms are parametrized in terms of the hidden 1-forms $B_{ab}, B_{a_1\cdots a_5}, \eta$.
In particular, we would like to investigate the conditions under which the gauge invariance of the FDA is realized once $A^{(3)}$ is expressed in terms of hidden 1-forms.
It is useful to first recall shortly the standard procedure for the  construction of a minimal FDA \footnote{A minimal FDA is one where the differential of any $p$-form does not contain forms of degree greater than $p$.} starting from an ordinary (super)Lie Algebra.

Let us denote by $\sigma^\Lambda$ the Maurer-Cartan 1-forms of the Lie algebra, and let us construct the so-called $(p+1)$-cochains $\Omega^{i\vert(p+1)}$ in some representation $D^i_j$ of the Lie group, that is $(p+1)$-forms of the type:
\begin{equation}\label{coch}
   \Omega^{i|(p+1)}=\Omega^i_{\Lambda_1\dots \Lambda_{p+1}}\sigma^{\Lambda_1}\wedge \dots \wedge \sigma^{\Lambda_{p+1}}
\end{equation}
where $ \Omega^i_{\Lambda_1\dots \Lambda_{p+1}}$ is a constant tensor. If the given cochains are cocycles, that is if they are closed, but not exact, they are elements of the Chevalley-Eilenberg  (CE in the following) Lie algebra cohomology.

When this happens, we can introduce a $p$-form $A^{i\vert (p)}$ and write the following new closed equation:
\begin{equation}\label{enlarge}
 d\,A^{i\vert(p)}+\Omega^{i\vert ({p+1})}=0
\end{equation}
which, together with the Maurer-Cartan equation of the Lie Algebra, is the first germ of a FDA, containing, besides the $\sigma^\Lambda$, also the new $p$-form $A^{i\vert(p)}$.

The procedure  can be now iterated taking as basis of new cochains $\Omega^{j\vert (p'+1)}$ the full set of forms, namely $\sigma^{\Lambda_i}$ and $\,A^{(p)}$, and look again for  cocycles.  If a new cocycle $\Omega^{j\vert (p'+1)}$ exists, then we can add again to the FDA a new equation
\begin{equation}\label{enlarge1}
 d\,A^{(p')}+\Omega^{j\vert (p'+1)}=0\,.
\end{equation}
The procedure can again be iterated till
no more cocycles can be found, obtaining in this way the largest FDA associated with the initial Lie algebra.
The extension of this  procedure to Lie superalgebras is  straightforward. Actually, in the supersymmetric case a set of non-trivial cocycles is generally present in superspace, due to the existence of Fierz identities obeyed by the wedge products of gravitino 1-forms.
In the case of supersymmetric theories, the 1-form fields of the superalgebra one starts with  are the vielbein $V^a$, the gravitino $\Psi$, the spin connection $\omega^{ab}$ and possibly a set of gauge fields.
However one should further impose the physical request that the FDA should be described in term of fields living
 in  ordinary \emph{superspace}, whose cotangent space is spanned by the supervielbein $\{V^a,\Psi\}$, dual to supertranslations. This corresponds to the
  physical request that the super Lie algebra has a fiber bundle structure, whose base space is spanned by the supervielbein, the rest of the fields spanning a fiber $\mathcal H$. This in turn implies an horizontality condition on the FDA, corresponding to gauge invariance:  the gauge fields belonging to $\mathcal{H} $ must be excluded from the construction of the cochains. In geometrical terms, this corresponds  to require that the CE-cohomology be restricted to  the so-called \emph{$\mathcal{H}$-relative CE-cohomology}.

 In the case of $D=11$ supergravity, one easily recognizes that the first step of the construction outlined above is  the introduction of the $\mathcal{H}$-relative 4-cocycle $\frac{1}{2}\overline{\Psi}\wedge \Gamma_{ab}\Psi \wedge V^a \wedge V^b $, which allows to define the 3-form $A^{(3)}$ of the FDA satisfying
 \begin{equation}
 dA^{(3)}=\frac 12 \overline{\Psi}\wedge \Gamma_{ab}\Psi \wedge V^a \wedge V^b\,,\label{coc1}
 \end{equation}
 that is eq. (\ref{FDA11a3}).
Including the new 3-form  $A^{(3)}$ in the basis of the relative cohomology of the supersymmetric FDA, we can perform the second step and construct a new cocycle of order seven, $ 15 A^{(3)}\wedge  dA^{(3)} +\frac{\ii}{2}\overline{\Psi}\wedge \Gamma_{a_1\cdots a_5}\Psi \wedge V^{a_1} \wedge \cdots V^{a_5}$, allowing the introduction of the 6-form $B^{(6)}$, satisfying:
\begin{equation}
 dB^{(6)}=15 A^{(3)}\wedge  dA^{(3)} +\frac{\ii}{2}\overline{\Psi}\wedge \Gamma_{a_1\cdots a_5}\Psi \wedge V^{a_1} \wedge \cdots V^{a_5}\,, \label{coc2}
 \end{equation}
that is eq.  (\ref{FDA11b6}).
The fact that the two cochains (\ref{coc1}) and(\ref{coc2}) are indeed cocycles is due to  Fierz identities in $D=11$, as reported in Appendix \ref{fierz}.

The second step defined above requires to enlarge the CE-relative cohomolgy to include the 3-form $A^{(3)}$.
We further remark that the inclusion of a new p-form, which is a gauge potential enjoying a gauge freedom, in the basis of the $\mathcal{H}$-relative CE-cohomology of the FDA, is physically meaningful only if  the whole of the FDA is gauge invariant. This in particular requires that the non-physical degrees of freedom in $A^{(3)}$ and $B^{(6)}$ are projected out from the FDA.

Let us turn now to the supersymmetric FDA of D=11 supergravity,  parametrized in terms of 1-forms. Now the symmetry structure  is based on the  hidden supergroup manifold $G$ which extends the super-Poincar\'e Lie group to include the extra hidden directions associated with the higher $p$-forms.
We note that the procedure  introduced in \cite{D'Auria:1982nx} and reviewed in Section \ref{11D} (see also \cite{Castellani:1991et}) can be thought of as the reverse of the costruction of a FDA from a given Lie superalgebra just recalled. Indeed, one starts from the physical FDA as given \emph{a priori} and tries to reconstruct, using the procedure of \cite{D'Auria:1982nx}, the hidden Lie superalgebra $\mathbb{G}$ that could have originated it using the algorithm of the CE-cohomology just described.

The hidden supergroup $G$ has the structure of a principal fiber bundle $(G/\mathcal{H},\mathcal{H})$, where $G/\mathcal{H}$ corresponds to superspace, the fiber $\mathcal{H}$ now including, besides the Lorentz transformations, also the hidden generators. More explicitly,
 let us rewrite the hidden Lie superalgebra $\mathbb{G}$ of $G$ as $\mathbb{G}={\mathcal{H}}+\mathbb{K}$, and decompose $\mathcal{H}=H_0 +H_b + H_f$, so that the generators $T_\Lambda \in \mathbb{G}$ are grouped
into $\{J_{ab}\}\in H_0$, $\{Z^{ab}; Z^{a_1\cdots a_5}\}\in H_b$, $\{Q'\} \in H_f$ and  $\{P_{a}; Q \}\in \mathbb{K}$. \footnote{Here and in the following with an abuse of notation we will use, for the cotangent  space of the group manifold $G$, spanned by the 1-forms $\sigma^\Lambda$,  the same symbols defined above   for the tangent space of $G$, spanned by the vector fields $T_\Lambda$.} We note that the subalgebra $H_b+H_f$ defines an abelian ideal of $\mathbb{G}$.

 The physical condition that the CE-cohomology be restricted to  the   $\mathcal{H}$-relative CE-cohomology corresponds now to the request that the FDA  be described in term of 1-form fields living on $G/\mathcal{H}$, and this in turn implies that the hidden 1-forms in $H_b$ and $H_f$, necessary for the parametrization of $A^{(3)}$ in terms of 1-forms, do not appear in $dA^{(3)}$ (see eq. (\ref{coc1})). Actually, as we shall see, the presence of the spinor 1-form $\eta$ is exactly what makes it possible to express $dA^{(3)}$ in terms of the relative cohomology only, that is in terms of the supervielbein.

\subsection{Gauge transformations from the hidden supergroup manifold}\label{BRS}
Taking into account the discussion above,
we now consider in detail the relation between the FDA gauge transformations and those of its hidden supergroup $G$\,.
The supersymmetric FDA, given in eq.s (\ref{FDA11omega}) - (\ref{FDA11b6}),
is left invariant under the  gauge transformations
\begin{eqnarray}\label{gauge11}
\left\{\begin{tabular}{l}
        $\delta A^{(3)}=d\Lambda^{(2)} $\\
         $\delta B^{(6)}=d\Lambda^{(5)} + \frac{15}2
         \, \Lambda^{(2)}\wedge  \overline\Psi \wedge \Gamma_{ab}\Psi \wedge V^a\wedge  V^b $
       \end{tabular}\right.
\end{eqnarray}
generated by the arbitrary forms $\Lambda^{(2)}$ and $\Lambda^{(5)}$.

The bosonic hidden  1-forms in $H_b$ are abelian gauge fields, whose gauge transformations are:
\begin{equation}\label{gauge}
\left\{
\begin{array}{l}
\delta_{b} B_{ab} = d\Lambda_{ab}\, ,\\
\delta_b B_{a_1\cdots a_5} = d \Lambda_{a_1\cdots a_5}\,
\end{array} \right.\,,
\end{equation}
 $\Lambda^{ab}$ and $\Lambda^{a_1\cdots a_5}$ being arbitrary Lorentz-valued scalar functions.

Requiring that $A^{(3)}$, parametrized in terms of 1-forms, transforms as (\ref{gauge11}) under the gauge transformations (\ref{gauge}) of the 1-forms, implies the gauge transformation of $\eta$ to be:
\begin{eqnarray}
\delta_b \eta = -E_2 \Lambda_{ab} \Gamma^{ab}\psi  - \ii E_3 \Lambda_{a_1\cdots a_5} \Gamma^{a_1\cdots a_5}\psi\,,\label{deltaeta}
\end{eqnarray}
consistently with the condition $D \delta \eta = \delta D \eta$.

 In this case the corresponding  $2$-form gauge parameter of $A^{(3)}$ turns out to be:
\begin{eqnarray}
\Lambda^{(2)} & = & T_0 \Lambda_{ab}   V^a \wedge V^b + 3 T_1 \Lambda_{a b}  B^{b} _{\;c}\wedge B^{c a}+ \nonumber\\ & +& T_2( 2 \Lambda_{b_1 a_1...a_4}  B^{b_1}_{\; b_2}\wedge B^{b_2 a_1...a_4} -B_{b_1 a_1...a_4} \Lambda^{b_1}_{\; b_2}\wedge B^{b_2 a_1...a_4})+ \nonumber\\ & +& 2T_3 \epsilon_{a_1...a_5 b_1...b_5 m}\Lambda^{a_1...a_5}\wedge B^{b_1...b_5}\wedge V^m +\nonumber \\
& +& 3 T_4 \epsilon_{m_1...m_6 n_1...n_5}\Lambda^{m_1m_2m_3p_1p_2}\wedge B^{m_4m_5m_6p_1p_2}\wedge B^{n_1...n_5} + \nonumber \\
& + &  S_2 \overline{\Psi}\wedge \Gamma^{ab}\eta   \Lambda_{ab}+ \ii S_3 \overline{\Psi}\wedge \Gamma^{a_1...a_5}\eta  \Lambda_{a_1...a_5}\,.\label{a3contr}
\end{eqnarray}

Considering also the gauge transformation  of the spinor 1-form  $\eta$  generated by the tangent vector in  $H_f$,  we have
\begin{equation}
\delta \eta = D\varepsilon ' +\delta_b \eta
\end{equation}
where we have introduced the infinitesimal spinor parameter $\varepsilon'$. The  2-form gauge parameter $\tilde\Lambda^{(2)}$ corresponding to the transformation in $H_f$  is then :
\begin{eqnarray}
\tilde\Lambda^{(2)} & = &
-\ii S_1 \overline{\Psi}\wedge \Gamma_{a} \varepsilon '  V^a - S_2 \overline{\Psi}\wedge \Gamma^{ab} \varepsilon '  B_{ab}- \ii S_3 \overline{\Psi}\wedge \Gamma^{a_1...a_5} \varepsilon '  B_{a_1...a_5}\,.\label{a3contr'}
\end{eqnarray}

In the following we are going to show that   all the
 diffeomorfisms in the hidden supergroup $G$, generated by Lie derivatives, are invariances of the FDA, the ones in the fiber $\mathcal{H}$ directions being  associated with a particular form of the gauge parameters of the FDA gauge transformations (\ref{gauge11}).

Let us first show that eq. (\ref{a3contr}) can be rewritten in a rather simple way using the contraction operator in the hidden Lie superalgebra $\mathbb{G}$ of $G$.
Defining the tangent vector:
\begin{equation}
{\vec z} \equiv \Lambda_{ab}Z^{ab}+\Lambda_{a_1\cdots a_5} Z^{a_1\cdots a_5}\in H_b\,,
\end{equation}
one finds that a gauge transformation leaving invariant  the $D=11$ FDA is recovered, once  $A^{(3)}$ is parametrized  in terms of 1-forms, if:
\begin{equation}
\Lambda^{(2)}=  \imath_{\vec z}(A^{(3)})\,,\label{assumpta2}
\end{equation}
where $\imath$ denotes the contraction operator. This result is actually true as a consequence of the set of relations
 (\ref{cond11}) obeyed by the coefficients of the parametrization (\ref{a3par}), that is under the same conditions required by supersymmetry for the consistency of the parametrization (\ref{a3par}).
Introducing the Lie derivative $\ell_{\vec z}\equiv d \imath_{\vec z} + \imath_{\vec z} d $,
we find the corresponding gauge transformation of $A^{(3)}$ to be:
\begin{equation}
\delta A^{(3)}= d \left(\imath_{\vec z}(A^{(3)})\right)= \ell_{\vec z} A^{(3)}\,.\label{gauge2}
\end{equation}
The last equality follows since $dA^{(3)}$, as given in (\ref{FDA11a3}), is invariant under transformations generated by $\vec z$ corresponding to the  gauge invariance of the supervielbein. Note that this is in agreement with the fact that the right hand side of $dA^{(3)}$ is in the relative $\mathcal H$ CE cohomology\,.

To recover the general gauge transformation of $B^{(6)}$ in terms of the hidden algebra would require the knowledge of its explicit parametrization  in terms of 1-forms, which at the moment we ignore. \footnote{Work is in progress on this topic.}
However, if we assume that its behavior under gauge transformations  be still generated by $\vec z$ through Lie derivatives, just like for $A^{(3)}$, namely if we require:
\begin{equation}
\Lambda^{(5)}=  \imath_{\vec z}(B^{(6)})\,,\label{assumpta}
\end{equation}
where $B^{(6)}$ is intended as parametrized in terms of 1-forms in $\mathbb{G}$, then a straightforward computation gives:
\begin{eqnarray}
\delta B^{(6)}&=&  \ell_{\vec z} B^{(6)}= d \left(\imath_{\vec z}(B^{(6)})\right)+ \imath_{\vec z}\left(d B^{(6)}\right)=d \Lambda^{(5)}+\imath_{\vec z}\left(15 A^{(3)} \wedge dA^{(3)}\right)\nonumber\\
&=& d \Lambda^{(5)}+\ 15 \Lambda^{(2)} \wedge dA^{(3)}\,,\label{gauge5}
\end{eqnarray}
which indeed reproduces eq. (\ref{gauge11}). The assumption (\ref{assumpta}) is corroborated by the analogous  computation in the seven dimensional model considered in Section \ref{7D}. In that case we can use, together with that of  $B^{(3)}$, the explicit parametrization of the Hodge dual related $B^{(2)}$  appearing in the dimensional reduction of the eleven dimensional 6-form $B^{(6)}$. As we shall see  the  assumption (\ref{assumpta}) can be fully justified if we think of $B^{(2)}$ as a remnant of $B^{(6)}$ in the dimensional reduction.

We stress that the gauge transformations (\ref{gauge2}) and (\ref{gauge5}) are not fully general, since the corresponding gauge parameters are not fully general, since they restricted to the ones satisfying (\ref{assumpta2}), (\ref{assumpta}).

\vskip 5mm
We should further still consider the  gauge transformations generated by the other elements of $\mathcal H$. Since the Lorentz transformations, belonging to $H_0 \subset \mathcal H$, are not effective on the FDA, all the higher $p$-forms being Lorentz-invariant, this analysis reduces to consider the transformations induced by the tangent vector $Q'\in H_f\subset \mathcal H$. Let us then consider:
\begin{equation}
 \vec q \equiv \bar \varepsilon' Q' \in H_f\,. \label{qvec}
\end{equation}
 We find $\delta_{\vec q}\eta=D\varepsilon '=\ell_{\vec q}\eta$ and:
\begin{eqnarray}
\delta_{\vec q}A^{(3)}
&=&- \ii S_1 \overline{\Psi}\wedge \Gamma_{a}D\varepsilon' V^a - S_2\overline{\Psi}\wedge \Gamma^{ab}D\varepsilon'   B_{ab} -\ii S_3\overline{\Psi}\wedge \Gamma^{a_1\cdots a_5}D\varepsilon'   B_{a_1\cdots a_5}\,\nonumber\\
&=&  d\imath_{\vec q} A^{(3)}=\ell_{\vec q}A^{(3)}\end{eqnarray}
where in the second line, after integration by parts, we used the relation on the $S_i$:
\begin{equation}
 S_1  +10 S_2-720S_3=0
 \end{equation}
following from 3-gravitino Fierz identities (see Appendix \ref{fierz}).

\subsection{The role of the nilpotent fermionic generator $Q'$}\label{q'}

In deriving the gauge transformations leaving invariant the supersymmetric FDA, in terms of hidden 1-forms, a crucial role is played by the spinor 1-form $\eta$ dual to the nilpotent generator $Q'\in H_f$. Indeed, besides the fact that it is required for the closure of the hidden superalgebra $\mathbb{G}$, it  also
guarantees the gauge invariance of the FDA, because of its non trivial gauge transformation, given in eq. (\ref{deltaeta}).

Actually, we may think of the spinor 1-form $\eta$ as playing the role of an intertwining field between the base superspace  and the fiber $\mathcal H$ of the principal fiber bundle corresponding to the hidden supergroup manifold $G=\{G/\mathcal H, \mathcal H\}$. This is also  evident from its covariant differential $D\eta$, eq. (\ref{Deta}), which is parametrized not only in terms of the supervielbein, as it happens for all the fields of the FDA and for $DB_{ab}$ and $DB_{a_1\cdots a_5}$, eq.(\ref{also}), but also in terms of the gauge fields in $H_b$, see eq. (\ref{deta}).
 In the following, we are going to clarify the role of $\eta$ in the more general context of the construction of FDAs discussed above, showing that its presence is essential to have a well defined, \emph{gauge invariant supersymmetric FDA}.

A clarifying example corresponds to considering a singular limit where $\eta$ is set equal to zero, so that its dual generator $Q'$ can be dropped out from $\mathbb{G}$. This limit may be obtained, in its simplest form, by redefining  the coefficients (\ref{cond11}) appearing in the parametrization of $A^{(3)}$ as follows:
\begin{equation}
E_2 \to E'_2=\epsilon E_2\,,\quad  E_3 \to E'_3=\epsilon^2 E_3\,,
\end{equation}
and   then taking the limit $\epsilon \to 0$.  One finds:
\begin{equation}
T_0\to \tilde T_0=\frac 16\,,\quad T_1 \to \tilde T_1= -\frac 1{90}\,,\quad T_2=T_3=T_4\to 0\,,\quad E_1=E_2=E_3\to 0\,,
\end{equation}
while $S_1,S_2,S_3 \to \infty$ in the limit. Recalling the parametrization of $A^{(3)}$, (\ref{a3par}), we see that setting $\eta=0$, the following finite limit can be obtained for $A^{(3)}$:
\begin{eqnarray}
A^{(3)} & \to & A^{(3)}_{lim} = \tilde T_0 B_{ab} \wedge V^a \wedge V^b +\tilde T_1 B_{a b}\wedge B^{b}_{\;c}\wedge B^{c a}\,.\label{a3lim}
\end{eqnarray}
so that its differential gives:
\begin{eqnarray}
dA^{(3)}_{lim} =\tilde T_0\left( \frac 1{2} \bar\Psi\Gamma_{ab}\Psi \wedge V^a \wedge V^b -   \ii   B_{ab} \wedge \bar\Psi\Gamma^{a}\Psi \wedge V^b\right) +\frac 3{2}\tilde T_1 \bar\Psi\Gamma_{ab} \wedge\Psi\wedge B^{b}_{\;c}\wedge B^{c a}\,.
\end{eqnarray}
We see that the parametrization (\ref{a3lim}) does not reproduce the FDA (\ref{FDA11a3}),being in fact obtained by a singular limit.  However this different FDA is based on the same hidden algebra $\mathbb{G}$, where now the cocycles are in the $H_0$-relative CE cohomology. Indeed   $dA^{(3)}_{lim}$ is now expanded on a basis of the \emph{enlarged superspace} $K_{enlarged}= K+H_b$, which includes, besides the supervielbein, also the bosonic hidden 1-forms. The  case where all the $E_i$ are proportional to the same power of $\epsilon$ can be done on the same lines, it again requires $\eta =0$ and leads to an $A^{(3)}_{lim}$ with all $\tilde T_i\neq 0$ (for $i=0,1,\cdots 4$). In this case $dA^{(3)}_{lim}$  is  expanded on a basis of the $K_{enlarged}$ also including $B_{a_1\cdots a_5}$.

A singular limit of the parametrization of $A^{(3)}$ was already considered in \cite{Bandos:2004xw}. The
limit considered in \cite{Bandos:2004xw} is similar to ours (where our parameters $E_i$   play a role similar to their parameter $s$)
\footnote{More precisely, the singular limit considered in \cite{Bandos:2004xw} is given in terms of a parmater $s\to 0$. The relation between their and our parameters is $s\propto E_2 -60 E_3$. }. There, the authors were studying the description of the hidden superalgebra as an expansion of $OSp(1|32)$. They observed  that a singular limit exists (which includes ours as a special case) such that  the  authomorphism group of the FDA is enlarged from what we called $\mathcal H$ to $Sp(32)$, but where the trivialization of the FDA in terms of an explicit $A^{(3)}$, written in terms of 1-forms, breaks down. From the above analysis we see that, at least for the restriction of the limit considered here, what does break down is  in fact  the trivialization of  the FDA on  \emph{ordinary superspace}, while a trivialization on $K_{enlarged}$ is still possible.

Note, however, that in this  case the gauge invariance of the new FDA requires that $B_{ab}$ (and analogously $B_{a_1 \cdots a_5}$) is not a gauge field anymore. Correspondingly,  $A^{(3)}_{lim}$ does not enjoy gauge freedom, all of its degrees of freedom  propagating in $K_{enlarged}$. It may then be interpreted  as a gauge-fixed form of $A^{(3)}$. Indeed,  \emph{it is precisely the gauge transformation of $\eta$, given in eq. (\ref{deltaeta}), that guarantees the gauge transformation of $A^{(3)}$ to be (\ref{gauge11})}. Actually, this relies on the fact that  $D\eta \in K_{enlarged}$ as we already observed previously, when we introduced eq. (\ref{deltaeta}).
Note that the transformation (\ref{gauge}), even if it is not  a gauge transformation in this limit case, still generates a diffeomorphism leaving invariant the new FDA (which is indeed based on the same supergroup $G$), since

\begin{equation}
\delta_{\vec z} A^{(3)}_{lim}= \ell_{\vec z} A^{(3)}_{lim}\,. \end{equation}
A gauge transformation bringing $A^{(3)}$ to $A^{(3)}_{lim}$ and, more generally, a gauge transformation such that $\eta' = \eta +\delta \eta =0$, is associated with transformations  generated by the tangent vector $\vec q$ introduced in (\ref{qvec}), in the particular case $\delta_{\vec q}\eta = D\varepsilon '= -\eta$.

In conclusion, the role of the extra fermionic nilpotent generator  amounts to require  the hidden  1-forms of the Lie superalgebra to be true gauge fields living on  the fiber $\mathcal{H}$ of the associated principal fiber bundle $\{G/\mathcal{H},\mathcal{H}\}$.\footnote{Note that this is equivalent to require that the construction of the FDA from Lie algebra of the supergroup $G$ be done using the $\mathcal H$-relative CE cohomology of $\mathbb{G}$.} It plays a role similar to a BRST ghost, since it guarantees that only the \emph{physical degrees of freedom} of the exterior forms appear in the supersymmetric FDA in a ``dynamical''  way: this amounts to say that, once the superspace is enlarged to $K_{enlarged}$, in the presence of $\eta$ and more generally of a non empty $H_f$, no explicit constraint has to be imposed on the fields, since the non-physical degrees of freedom of the fields in $H_b$ and in $H_f$ transform into each other and do not contribute to the FDA.

\section{The hidden gauge algebra of $D=7$, $\mathcal{N}=2$ Free Differential Algebra}\label{7D}

The same procedure explained in the eleven dimensional case can be applied to lower dimensional supergravity theories, in order to associate to any such theory containing $p$-forms  (with $p>1$),  a hidden  Lie superalgebra containing, as a subalgebra, the super-Poincar\'e algebra.
Since in the D=11 theory the closure of the FDA and of the corresponding hidden superalgebra are strictly related to 3-gravitino Fierz identities of the given theory, the same must happen in any lower dimensions.

As an  interesting example we consider in this section the minimal $D=7$, $\mathcal{N}=2$ theory (not coupled to matter), where the hidden structure turns out to be particularly rich since, as we will see, in its most general form it includes two nilpotent fermionic generators.

Working as in the eleven dimensional case within the geometric formulation of superspace $p$-forms, its physical content on space-time is given  by the vielbein 1-form $V^a$, a triplet of vectors 1-forms $A^x$ ($x=1,2,3$), a 2-form $B^{(2)}$, together with a  gravitino 1-form Dirac spinor which we describe as a couple of 8-component spin-3/2 pseudo-Majorana fields $\psi_{A\mu}$ ($A=1,2$)  satisfying the reality condition $\overline\psi^A = \epsilon^{AB} (\psi_B)^T$. \footnote{The charge conjugation matrix in D=7 can always be chosen $C=\bfone$.}

The interacting $D=7$ minimal theory  was studied, at the lagrangian level, by many authors \cite{Townsend:1983kk}-\cite{Fre':2015lds}. In particular, in \cite{Townsend:1983kk} it was observed that one can trade the 2-form formulation of the theory by a formulation in terms of a  3-form, $B^{(3)}$, the two being related by Hodge-duality of the corresponding field strengths on space-time, and they give rise to different lagrangians.
From our point of view, where the FDA is considered (and not a Lagrangian description), both forms are required for a fully general formulation, together with a triplet of 4-forms, $A^{x\vert(4)}$, whose field strengths are Hodge-dual to the gauge vectors $A^x$.

One of the main reasons for choosing the minimal $D=7$ model is related to the fact that in this case  we will be able to find an explicit parametrization in terms of 1-forms of both $B^{(2)}$ and $B^{(3)}$, whose field strengths are related by Hodge duality. We will find that in this case a general parametrization requires  the presence of two independent hidden spinor  1-forms. Since $B^{(2)}$ in $D=7$ can be obtained by dimensional reduction of $B^{(6)}$ in the eleven dimensional FDA, this investigation also allows to shed some light on the extension of the hidden superalgebra of $D=11$ supergravity when also the parametrization of $B^{(6)}$, still unknown, would be considered (see section \ref{include3}).

The minimal $N=2, D=7$ supergravity is based on the following  supersymmetric FDA:
\begin{align}
& R^{ab}\equiv d\omega^{ab}-\omega^{a}{}_c\wedge \omega^{cb}=0 \,,\label{fdaomega}\\
& T^a\equiv D V^a -\frac{\ii}{2}\overline{\psi}^A \wedge \Gamma^a \psi_A =0\,,\label{fdav} \\
& \rho \equiv D \psi =0\,,\label{fdapsi}\\
& F^x \equiv d {A}^x - \frac{\ii}{2} \sigma^{x \vert B}_{\;\;\;\;\;A} \overline{\psi}^A\wedge \psi_B=0\,,\label{fdaa1} \\
& F^{(3)}\equiv  d {B}^{(2)}+d {A}^x\wedge {A}^x - \frac{\ii}{2} \overline{\psi}^A \wedge \Gamma_a \psi_A \wedge V^a =0\, ,\label{fdab2} \\
& G^{(4)}\equiv d {B}^{(3)}-\frac{1}{2} \overline{\psi}^A \wedge \Gamma_{ab}\psi_A \wedge V^a \wedge V^b=0\, , \label{fdab3}\\
& F^{x(4)}\equiv d{A}^{x\vert (4)}  + \frac{1}{2}\left(d{A}^x \wedge  {B}^{(3)} - {A}^x \wedge d {B}^{(3)}\right)-\frac{1}{6} \sigma^{x \vert B}_{\;\;\;\;\;A} \overline{\psi}^A \wedge \Gamma_{abc}\psi_B \wedge V^a \wedge V^b \wedge V^c =0\,, \label{fdaa4}
\end{align}
where now $D$ denotes the $D=7$ Lorentz-covariant differential and $\sigma^{x \vert B}_{\;\;\;\;\;A}$ are the usual Pauli matrices. As already mentioned the $d^2$-closure of this FDA relies on the Fierz identities relating gravitino 3- and 4-forms currents in $D=7$.

To find the hidden superalgebra, let us introduce the following set of bosonic Lorentz-indexed 1-forms: $B_a$, associated with $B^{(2)}$, $B_{ab}$, associated with $B^{(3)}$, $A^x_{abc}$, associated with $A^x_{(4)}$, requiring  their Maurer-Cartan equations  to be:
\begin{align} \label{d1form}
& D B_{ab} = \alpha \overline{\psi}^A \wedge \Gamma_{ab}\psi_A ,  \nonumber\\
& D B_a = \beta \overline{\psi}^A \wedge \Gamma_a \psi_A , \nonumber\\
& D {A}^{x\vert}_{\;\;\; abc} = \gamma \sigma^{x \vert B}_{\;\;\;\;\;A} \overline{\psi}^A \wedge \Gamma_{abc}\psi_B \,.
\end{align}
whose integrability conditions are automatically satisfied since $R^{ab}=0$.
The arbitrary choice of the coefficients in the right-hand-side fixes the normalization of the bosonic 1-forms $B_a$, $B_{ab}$ and ${A}^{x\vert}_{\;\;\; abc}$. In the following, we will choose $\alpha=\frac 12$, $\beta=\frac{\ii}2$, $\gamma=\frac16$.

The bosonic forms $B^{(2)}$ and $B^{(3)}$ will be parametrized, besides the 1-forms $V^a,A^x$ already present in the FDA, also in terms of the new 1-forms $B_a, B_{ab}, A^{x\vert}_{\;\;\; abc}$, and as we are going to show, the consistency of their parametrizations   also requires the presence of \emph{two} nilpotent fermionic 1-forms, $\eta_A$ in the parametrization of $B^{(2)}$ and $\xi_A$ in the one of $B^{(3)}$, whose covariant derivatives satisfy:
\begin{align}
& D \eta_A = l_1 \Gamma_a \psi_A \wedge V^a + l_2  \Gamma^a \psi_A \wedge B_a + l_3 \Gamma^{ab}\psi_A \wedge B_{ab} + \nonumber \\
& + l_4 \psi_B \sigma^{x \vert B}_{\;\;\;\;\;A} \wedge {A}^x + l_5 \Gamma^{abc}\psi_B \sigma^{x \vert B}_{\;\;\;\;\;A} \wedge {A}^{x\vert}_{\;\;\; abc},\label{deta7}\\
& D \xi_A = e_1 \Gamma_a \psi_A \wedge V^a + e_2 \Gamma^a \psi_A \wedge B_a + e_3 \Gamma^{ab}\psi_A \wedge B_{ab} + \nonumber \\
& + e_4 \psi_B \sigma^{x \vert B}_{\;\;\;\;\;A} \wedge {A}^x + e_5 \Gamma^{abc}\psi_B \sigma^{x \vert B}_{\;\;\;\;\;A} \wedge {A}^{x\vert}_{\;\;\; abc}\,,\label{dxi}
\end{align}
where $l_i$, $e_i$ are so far unspecified structure constants of the hidden superalgebra, constrained to satisfy (from the integrability of $D\eta_A$ and $D\xi_A$ and use of the Fierz identities):
\begin{align}
& - \ii l_1 -  \ii l_2  + 6 l_3  - \ii l_4 -10 l_5  = 0, \label{inteta} \\
& - \ii e_1-\ii e_2 +6 e_3  - \ii e_4 -10 e_5  =0 .\label{intxi}
\end{align}
The consistency of the parametrizations amounts to require that the differential of $B^{(2)}$ and $B^{(3)}$, as given in equations (\ref{fdab2}) and  (\ref{fdab3}), must be reproduced by the differential of their parametrizations (\ref{B2par}), (\ref{B3par}).
This is  analogous to what happens in $D=11$; in that case, however, only the parametrization of the 3-form was considered, and its closure, besides the precise values of the coefficients, required the presence of just \emph{one} spinor 1-form dual to a nilpotent fermionic generator.

Explicitly we give the following general Ansatz for the parametrization of $B^{(2)}$ and $B^{(3)}$  in terms of the $1$-forms $\{V^a,\psi_A,B_a,B_{ab},{A}^{x \vert}_{\;\;\; abc},\xi_A,\eta_A\} $\footnote{We should in principle also consider the parametrization of the 4-form $A^x_{(4)}$. This  deserves further investigation. Some work is in progress on this point.}:
\begin{align}
 {B}^{(2)} =& \sigma B_a \wedge V^a + \tau \overline{\psi}^A \wedge \eta_A ,\label{B2par}
\\
 {B}^{(3)} = & \; \tau_0 B_{ab}\wedge V^a \wedge V^b + \tau_1  B_{ab}\wedge B^a V^b + \tau_2  B_{ab}\wedge B^a B^b +  \tau_3 B_{ab}\wedge B^{bc}\wedge B_c^{\; a}+ \nonumber \\
& +   \epsilon_{ab_1...b_3c_1...c_3}( \tau_4\,V^a + \tau_5 \,B^a)\wedge {A}^{x \vert b_1...b_3} \wedge {A}^x_{\; c_1...c_3}+ \nonumber \\
& + \tau_6  B_{ab}\wedge {A}^x_{\; acd}\wedge {A}^{x\vert bcd}+ \tau_7  \epsilon_{xyz} {A}^x \wedge A^y_{\; abc}\wedge {A}^{z\vert abc} + \nonumber \\
& + \tau_8 \epsilon_{xyz} {A}^x \wedge A^y \wedge A^z +  \tau_9 \epsilon_{xyz} \epsilon_{abcdlmn}A^{x\vert abc}\wedge A^{y \vert dlp}\wedge A^{z \vert mn}_{\;\;\;\;\;\;\;\;p}+ \nonumber \\
& + \sigma_1 \overline{\psi}^A \wedge \Gamma_a \xi_A \wedge V^a + \sigma_2  \overline{\psi}^A \wedge \Gamma_a \xi_A \wedge B^a +  \sigma_3 \overline{\psi}^A  \wedge \Gamma_{ab}\xi_A \wedge B^{ab}+ \nonumber \\
& +\sigma_4 \overline{\psi}^A \wedge \xi_B \sigma^{x \vert B}_{\;\;\;\;\;A} \wedge {A}^x +  \sigma_5 \overline{\psi}^A \wedge \Gamma^{abc}\xi_B \sigma^{x \vert B}_{\;\;\;\;\;A} \wedge {A}^{x \vert}_{\;\;\; abc}\,.\label{B3par}
\end{align}

The set of coefficients $\{\tau_j\}$, $\{\sigma_i\}$ are determined by requiring that the parametrizations (\ref{B2par}) and (\ref{B3par}) satisfy the FDA, in particular eq.s (\ref{fdab2}), (\ref{fdab3}). Their explicit  expression is given in Appendix \ref{coeff}.
However, we still have the freedom to fix the normalization of the spinor 1-forms $\xi_A$, $\eta_A$. We are going to fix them in order to obtain a simple expression. In particular we choose the normalization of $\eta_A$ by imposing, in the parametrization of $B^{(2)}$,
$\tau=1$.
As far as the normalization of $\xi_A$  is concerned using  the general solution for the coefficients given in Appendix \ref{coeff},  we find $\frac{e_2}{\sigma_2}=\frac{e_5}{\sigma_5}\equiv H$, where, with the normalization chosen for the bosonic 1-forms:
\begin{equation}
H=-2\left(e_1+e_2 -2\ii e_3\right)\left(e_1+e_2 -2\ii e_5\right)\,.
\end{equation}
Therefore we choose $H=1$, which is a valid normalization in all cases where $H\neq 0$, that is for $e_1+e_2 \neq 2\ii e_3$ or $e_1+e_2 \neq 2\ii e_5$. Actually the general solution  given in Appendix \ref{coeff} shows that to choose $\tau\neq 0$, $H\neq 0$ are not   restrictive assumptions, since the cases $\tau =0$ and/or  $H =0$  would correspond to  singular limits where the gauge structure of the supersymmetric FDA breaks down. This is strictly analogous to what we discussed in Section \ref{q'} for the D=11 case as far as the gauge structure of the theory is concerned.

With the above normalizations we obtain:

\begin{equation}
\sigma = 2 \ii l_2, \;\;\; l_1 = \frac{\ii}{2}\left(-1+2 \ii  l_2 \right), \;\;\; l_4 = \frac{\ii}{2},
\end{equation}
\begin{eqnarray}
&&\tau_0= 2 \left[\ii e_1 (e_3-e_5)+\left(\frac{\ii}{2}e_2+e_3\right)\left(\frac{\ii}{2}e_2+e_5\right)\right]\nonumber\\
&&\tau_1= -4\ii e_2(\ii e_2 + 2 e_5)\,,\quad \tau_2= -2 e_2^2\,,\quad \tau_3= -\frac 83 e_3(e_3-2e_5)\nonumber\\
&& \tau_4 = e_5(\ii e_2 +2 e_5)\,,\quad \tau_5 =-\ii e_2e_5\,,\quad \tau_6= 36 e_5^2\nonumber\\
&&\tau_7 = -12 e_5^2 \,,\quad \tau_8= \frac 23 e_4[e_1+e_1-6\ii(e_3+e_5)]\,,\quad \tau_9= -3e_5^2\nonumber\\
&&\sigma_1= -e_1-2e_2+4\ii e_5\,,\quad \sigma_2= e_2\,,\quad \sigma_3= -e_3+2e_5\,,\quad \sigma_4= -e_4\,,\quad \sigma_5=e_5\label{taus}\,,
\end{eqnarray}
where the $e_i$ are constrained by  (\ref{intxi}).
\footnote{We observe that the combination $\tau_4\,V^a + \tau_5 \,B_a\equiv \tilde B_a$ could be used, instead of $B_a$, in the parametrization of $B^{(3)}$. This redefinition  simplifies the expression of $B^{(3)}$, in particular the term $B_{ab}\wedge \tilde B^a\wedge V^b $ vanishes.
}

\subsection{The  hidden superalgebra}\label{hidden}

Let us write now, analogously to what was done in $D=11$, the $D=7$ hidden superalgebra in terms of generators $T_\Lambda$ dual to the set of 1-forms $\sigma^\Lambda$ of the theory. In this case we have
\begin{equation}
\sigma^\Lambda=\{V^a,\psi_A,\omega^{ab}, B_a,B_{ab},{A}^{x \vert}_{\;\;\; abc},\xi_A,\eta_A\}\label{sigma7d}
\end{equation}
and
\begin{equation}
T_\Lambda= \{P_a,Q_A,J_{ab},T^x , Z^a, Z^{ab},T^{x \vert abc},Q'_A,Q''_A\}\,.\label{t7d}
\end{equation}
 The only non-trivial mappings are now:
\begin{align}
& \psi_A (Q^B) = \delta _A ^{\; B}, \; \; \; \xi_A(Q'^B)=\delta_A^{\; B}, \; \; \; \eta_A(Q''^B)= \delta_A ^{\; B}, \nonumber \\
& V^a(P_b) = \delta^a _{\; b},  \; \; \; B_a(Z^b) = \delta^b _{\; a}, \; \; \; B_{ab}(Z^{cd}) = 2\delta^{cd} _{\; ab}, \nonumber \\
& {A}^x(T^y)= \delta^{xy}, \; \; \; {A}^{x \vert}_{\;\;\; abc}(T^{y\vert lmn})= 3! \delta^{xy}\delta_{abc}^{\; lmn}, \;\;\; B_{a_1\cdots a_5}(Z^{b_1\cdots b_5})= 5!\delta_{a_1\cdots a_5}^{b_1\cdots b_5} \,,
\end{align}
so that the (anti)-commutators of the superalgebra (besides the Poincar\'e Lie algebra) can be written as
\begin{align}
& \lbrace Q^A, \overline{Q}_B \rbrace = - \ii \Gamma^a \left( P_a+\eta_{ab}Z^b \right) \delta^A_{\; B} - \frac 12 \Gamma_{ab}Z^{ab}\delta^A_{\; B}-\sigma^{x \vert A}_{\;\;\;\;\; B}\left(\ii T^x +\frac 1{18} \Gamma^{abc}T^x_{\; abc}\right)\,, \\
& [Q_A, P_a]= -2 \Gamma_a (e_1   Q'_A+l_1Q''_A)\,, \\
& [Q_A, Z^a]= -2 \Gamma^a (e_2Q'_A+l_2Q''_A)\,, \\
& [Q_A, Z^{ab}]= -4  e_3 \Gamma^a Q'_A\,,\\
& [Q_A, T^x]= - 2 \sigma^{x \vert B}_{\;\;\;\;\;A}(e_4Q'_B+ l_4 Q''_B) \,, \\
& [Q_A, T^{x \vert abc}]=-12 e_5 \Gamma^{abc}\sigma^{x \vert B}_{\;\;\;\;\;A}Q'_B\,,\\
& [J_{ab}, Z^{c}]=-2 \delta^{c}_{[a}Z_{b]}\,,\\
& [J_{ab}, Z^{cd}]=-4\delta^{[c}_{[a}Z^{d]}_{b]}\,,\\
&[J_{ab}, T^{x | c_1c_2 c_3}]=-12 \delta^{[c_1}_{[a}T^{x \vert \;c_2  c_3]}_{\;\;\;\;b]}\,,\\
& [J_{ab}, Q]=- \Gamma_{ab} Q\,,\\
& [J_{ab}, Q']=- \Gamma_{ab} Q'\,.\end{align}
All the other possible commutators vanish.

\paragraph{Lagrangian subalgebras.\\}

Let us consider here two relevant subalgebras, that we call ``electric hidden subalgebras" or ``lagrangian subalgebras" because of their relevance for the construction of the lagrangian.
 The first one corresponds to consider the FDA restricted to only include as bosonic external forms  $A^x$ and $B^{(2)}$, but not their mutually non-local forms  $B^{(3)}$ and $A^x_{(4)}$, respectively. This is the appropriate framework for the construction of the lagrangian in terms of $B^{(2)}$, as considered for example in \cite{Salam:1983fa},\cite{Bergshoeff:1985mr}. In this case only the nilpotent spinor 1-form $\eta$, corresponding to the generator $Q''$, appears in the hidden subalgebra,  which is then  obtained  by the truncation $Q'_A \to 0$. Note that  the same subalgebra can be obtained equivalently if the full FDA is considered, but we take the truncation so that the same nilpotent spinor 1-form  appears in the parametrizations (\ref{B2par}) and (\ref{B3par}); namely $\eta_A=\xi_A$ and
 $Q'_A=Q''_A= \frac 12 \hat Q_A$.  In this case the
 Maurer-Cartan equations (\ref{deta7}) and (\ref{dxi}) coincide, implying $\{e_i\}= \{l_i\}$, so that in particular $e_3=e_5=0$, since $l_3=l_5=0$. This in turn implies, on the set of $\{\tau_j\}$ given in (\ref{taus}), that all the contributions in $B_{ab}$ and $A^x_{abc}$ in the parametrization of $B^{(3)}$ disappear, so that the corresponding generators $Z^{ab}$ and $T^{x \vert abc}$ decouple and can be set to zero. The resulting subalgebra is:
 \begin{align}
& \lbrace Q^A, \overline{Q}_B \rbrace = - \ii \Gamma^a \left( P_a+\eta_{ab}Z^b \right) \delta^A_{\; B} -\ii \sigma^{x \vert A}_{\;\;\;\;\; B}  T^x , \\
& [Q_A, P_a]= -2 e_1\Gamma_a    \hat Q_A, \\
& [Q_A, Z^a]= -2 e_2\Gamma^a \hat Q_A, \\
& [Q_A, T^x]= - 2 e_4\sigma^{x \vert B}_{\;\;\;\;\;A}\hat Q_B \,.
\end{align}

The alternative lagrangian subalgebra is found starting instead from the restricted FDA where $B^{(2)}$ is excluded, together with  $A^x_{(4)}$, which is the appropriate framework for the construction of the lagrangian  in terms of $B^{(3)}$ only \cite{Townsend:1983kk}.
In this case the spinor $\eta$ appearing in the parametrization of $B^{(2)}$ drops out together with its dual generator $Q''_A\to 0$. The 1-forms $B_a$ and $A^x_{abc}$ could still be included in the parametrization of $B^{(3)}$ as trivial deformations, and they can be consistently decoupled by setting $e_2=e_5=0$.

Let us stress that  both Lagrangian subalgebras require the truncation of the superalgebra to only one out of  the two nilpotent spinors.

\vskip 5mm

The analysis above shows that the full superalgebra hidden in the supersymmetric $D=7$ FDA is larger than the one just involving the fields appearing in the Lagrangian in terms of either $B^{(2)}$ or of $B^{(3)}$ only. This is in fact a well know fact, which holds  in four dimensional extended supersymmetric theories, where the central extension of the supersymmetry algebra is associated to electric and magnetic charges \cite{Witten:1978mh}, while the electric subalgebra only involves electric charges whose associated gauge potentials will appear in the lagrangian description.

The superalgebra we have constructed above includes all the dynamical fields of the FDA together with their Hodge-duals, and is in this sense fully general. It requires the presence of 2 independent nilpotent spinorial charges. Let us remark that the analysis above shows that two independent spinorial generators are necessary if we want to include in the hidden algebra involving $B^{(2)}$ and  $B^{(3)}$ also the field $A^x_{abc}$ associated to  $A^{x(4)}$. However, we did not consider in the above description the non-dynamical volume form $F^{(7)}= dB^{(6)}+ \cdots$, with its possible associated 1-form $B_{a_1\cdots a_5}$. We could ask if the inclusion of such extra contributions in the parametrization of $B^{(2)}$ and  $B^{(3)}$ could significantly alter the results obtained, and if it would requires the presence of extra spinorial charges. This issue is discussed in the following subsection.

\subsection{Including $B_{a_1\cdots a_5}$}
To complete the analysis of the minimal theory in $D=7$, let us further investigate  the superalgebra hidden in the extension of the FDA to include the (non-dynamical) form $B^{(6)}$ associated with the volume form in seven dimensions.

 It contributes to the FDA as:
 \begin{equation}
 dB^{(6)} - 15 B^{(3)}\wedge dB^{(3)} = \frac \ii 2 \bar\psi \wedge \Gamma_{a_1\cdots a_5} \psi \wedge V^{a_1} \cdots \wedge V^{a_5}\,,
 \end{equation}
 as it is evident by the dimensional  reduction of the eleven dimensional 6-form, that we will treat in Section \ref{11d7d}.
 The aim is double: on one hand we would like to check how the hidden algebra gets enlarged in the presence of the extra 1-form $B_{a_1\cdots a_5}$ associated with $B^{(6)}$, and in particular if it requires the presence of one more fermionic generator; on the other hand, this analysis will turn out to be useful once we will relate, in the next section, the D=7 theory to the D=11 one.

Let us quote below the result. We require the covariant derivative of the spinor 1-forms to be now:
\begin{align}
& D \xi_A = e_1 \Gamma_a \psi_A \wedge V^a + e_2 \Gamma^a \psi_A \wedge B_a + e_3 \Gamma^{ab}\psi_A \wedge B_{ab} + \nonumber \\
& + e_4 \psi_B \sigma^{x \vert B}_{\;\;\;\;\;A} \wedge {A}^x + e_5 \Gamma^{abc}\psi_B \sigma^{x \vert B}_{\;\;\;\;\;A} \wedge {A}^{x\vert}_{\;\;\; abc} + e_6 \Gamma^{a_1...a_5}\psi_A B_{a_1...a_5}\,,\label{include1}\\
& D \eta_A = l_1 \Gamma_a \psi_A \wedge V^a + l_2 \Gamma^a \psi_A \wedge B_a + l_3 \Gamma^{ab}\psi_A \wedge B_{ab} + \nonumber \\
& +l_4 \psi_B \sigma^{x \vert B}_{\;\;\;\;\;A} \wedge {A}^x + l_5 \Gamma^{abc}\psi_B \sigma^{x \vert B}_{\;\;\;\;\;A} \wedge {A}^{x\vert}_{\;\;\; abc} + l_6 \Gamma^{a_1...a_5}\psi_A B_{a_1...a_5}\label{include2}\,.
\end{align}
and, besides eq.s (\ref{d1form}), we define:
\begin{align}
& D B_{a_1...a_5} = \frac{\ii}{2} \overline{\psi}^A \wedge \Gamma_{a_1...a_5} \psi_A, \,.
\end{align}

The integrability conditions of (\ref{include1}) and (\ref{include2}) give:

\begin{align}
& - \ii l_1 -  \ii l_2  + 6 l_3  - \ii l_4 -10 l_5 -\ii 360 l_6 = 0, \label{inteta1} \\
& - \ii e_1-\ii e_2 +6 e_3  - \ii e_4 -10 e_5 -\ii 360 e_6  =0 .\label{intxi1}
\end{align}

We find the following new parametrizations for $B^{(2)}$ and $B^{(3)}$:

\begin{align}
&{B}^{(2)} = B^{(2)}_{old} + \chi \epsilon_{a_1...a_5 a b} B^{a_1...a_5}\wedge B^{ab}\\
&{B}^{(3)} = \; B^{(3)}_{old} + \tau_{10}B_{a a_1...a_4}\wedge B^a_{\; b}\wedge B^{b a_1...a_4}  + \tau_{11} \epsilon_{a_1...a_5 a b}B^{a_1...a_5}\wedge V^a \wedge V^b + \nonumber \\
& + \tau_{12} \epsilon_{a_1...a_5 a b}B^{a_1...a_5}\wedge B^a \wedge V^b  + \tau_{13} \epsilon_{a_1...a_5 a b}B^{a_1...a_5}\wedge B^a \wedge B^b + \nonumber \\
& + \tau_{14} \epsilon_{a_1...a_5 a b}B^{a_1...a_5}\wedge {A}^x_{\; acd}\wedge {A}^{x\vert bcd} +
  + \sigma_6 \overline{\psi}^A \wedge \Gamma_{a_1...a_5} \xi_A \wedge B^{a_1...a_5}\,.
\end{align}
where $ B^{(2)}_{old}$ and  $ B^{(3)}_{old}$ are given by equations (\ref{B2par}) and (\ref{B3par}).
The values of the new set of coefficients is given in Appendix \ref{coeff}.

The result is that the parametrization of the extended forms in terms of 1-forms is more complicated in this case, but the closure of the hidden superalgebra does not require any new spinorial 1-form generator besides $\xi_A$ and $\eta_A$.

To express the superalgebra  in the dual form, it is sufficient to introduce the bosonic generator $Z^{a_1\cdots a_5}$ satisfying $B_{a_1\cdots a_5}(Z^{b_1\cdots b_5})= 5!\delta_{a_1\cdots a_5}^{b_1\cdots b_5}$, and we get
\begin{align}
& \lbrace Q^A, \overline{Q}_B \rbrace = - \left[\ii \Gamma^a \left( P_a+\eta_{ab}Z^b \right) +  \frac {1}{2} \Gamma_{ab}Z^{ab}+
 \frac {\ii}{5!}\Gamma_{a_1\cdots a_5}Z^{a_1\cdots a_5}\right]\delta^A_{\; B}+\nonumber\\
 &\hskip 2.2cm -\sigma^{x \vert A}_{\;\;\;\;\; B}\left(\ii T^x +\frac 1{18} \Gamma_{abc}T^{x\vert \; abc}\right), \\
& [Q_A, P_a]= -2 \Gamma_a (e_1   Q'_A+l_1Q''_A), \\
& [Q_A, Z^a]= -2 \Gamma^a (e_2Q'_A+l_2Q''_A), \\
& [Q_A, Z^{ab}]=-4 \Gamma^{ab}(e_3  Q'_A + l_3 Q''_A), \\
& [Q_A, T^x]= - 2 \sigma^{x \vert B}_{\;\;\;\;\;A}(e_4Q'_B+ l_4 Q''_B) , \\
& [Q_A, T^{x \vert abc}]= -12 e_5 \Gamma^{abc}\sigma^{x \vert B}_{\;\;\;\;\;A}Q'_B\\
& [Q_A, Z^{a_1...a_5}]=-2 (5!) \Gamma^{a_1...a_5}(e_6  Q'_A + l_6 Q''_A),.
\end{align}

\subsection{Gauge structure of the minimal D=7 FDA}

The gauge structure of the $D=7$ FDA can be analyzed in a strict analogous way as we have done for the $D=11$ case.
We limit ourselves to give just a short discussion of it since the relevant point about the role of the nilpotent charges dual to the spinor $1$-forms $\eta_A$ and $\xi_A$ is completely analogous to the one discussed for $\eta$ in the $D=11$ case.
The supersymmetric FDA is invariant under the following  gauge transformations:
\begin{eqnarray} \label{gaugex}
\left\{ \begin{array}{l}
\delta A^x = d \Lambda^x \,, \\
\delta B^{(2)}= d\Lambda^{(1)}-\Lambda^x dA^x\, , \\
\delta B^{(3)} =d \Lambda^{(2)} \,, \\
\delta A^{x \vert (4)} =d \Lambda^{x \vert (3)} -\frac{1}{2}(\Lambda^x dB^{(3)}+\Lambda^{(2)}\wedge dA^x)\,,\\
\delta  B^{(6)} =d \Lambda^{(5)} -15 \Lambda^{(2)} \wedge dB^{(3)}\,.
\end{array} \right.
 \label{gauge7d}
\end{eqnarray}
Analogously to the eleven dimensional case, the gauge transformations (\ref{gauge7d}) leaving invariant the FDA can be obtained, for particular $(p-1)$-form parameters, through Lie derivatives acting on the hidden symmetry supergroup $G$ underlying the theory.
In this case,   $G$  has the fiber bundle structure $G= \mathcal{H} + K$, where now  $K=G/\mathcal{H}$ is spanned by the supervielbein $\{V^a,\psi_A\}$. The fiber $\mathcal{H}= H_0+ H_b+H_f$ is generated by the Lorentz generators in $H_0$ and by the gauge and hidden generators in $H_b$ and $H_f$, where now $\{T^x,Z^a,Z^{ab},T^{x|abc},Z^{a_1\cdots a_5}\}$ span  $H_b$, while $\{\xi_A,\eta_A\}$ span $H_f$.

Explicitly, let us define the tangent vector in $H_b$:
\begin{eqnarray}
\vec z \equiv \Lambda^x T^x + \Lambda_a Z^a + \Lambda_{ab} Z^{ab} +\Lambda^{x}_{abc} T^{x|abc} + \Lambda_{a_1\cdots a_5} Z^{a_1\cdots a_5}\in H_b
\end{eqnarray}
By straightforward calculation we get that the gauge transformations of  $A^x$, $B^{(2)}$ and $B^{(3)}$ in (\ref{gauge7d})  can be obtained by requiring:
\begin{eqnarray}
\delta A^x &=& \ell_{\vec z}A^x\,,\\
\delta B^{(2)}&=& \ell_{\vec z} B^{(2)}\,,\\
\delta B^{(3)}&=& \ell_{\vec z} B^{(3)}
\end{eqnarray}
for the choice of $(p-1)$-form gauge parameters:
\begin{eqnarray}
\Lambda^x&=& \imath_{\vec z} A^x\,,\\
\Lambda^{(1)}&=& \imath_{\vec z} B^{(2)}\,,\\
\Lambda^{(2)}&=& \imath_{\vec z} B^{(3)}\,,
\end{eqnarray}
provided  the values of the $\tau_i$ and $\sigma_i$ parameters be given by the equation (\ref{b3coe}) of Appendix \ref{coeff}, which also assure supersymmetry and consistency of the theory.
We expect that in general, also for the forms $A^{x \vert (4)}$ and $B^{(6)}$, whose parametrizations in terms of 1-forms are still unknown, the rest of the gauge transformations in (\ref{gauge7d}) leaving invariant the supersymmetric FDA should be:
\begin{eqnarray}
\delta A^{x|(4)} &=& \ell_{\vec z}A^{x|(4)}\,,\\
\delta B^{(6)}&=& \ell_{\vec z} B^{(6)}\,,
\end{eqnarray}
for the choice of $(p-1)$-form gauge parameters:
\begin{eqnarray}
\Lambda^{x|(3)}&=& \imath_{\vec z} A^{x|(4)}\,,\\
\Lambda^{(5)}&=& \imath_{\vec z} B^{(6)}\,.
\end{eqnarray}
This corresponds to the following gauge transformations of the 1-forms in $H_b$:
\begin{eqnarray}
\left\{ \begin{array}{l}
\delta A^x = d \Lambda^x \,, \\
\delta B_a= d\Lambda_a\, , \\
\delta B_{ab} =d \Lambda_{ab} \,, \\
\delta A^{x}_{abc} =d \Lambda^{x}_{abc}\,,\\
\delta  B_{a_1\cdots a_5} =d \Lambda_{a_1\cdots a_5}\,,
\end{array} \right.
\label{gauge7hb}
\end{eqnarray}
together with the gauge transformations of the 1-forms in $H_f$:
\begin{eqnarray}
\left\{ \begin{array}{l}
\delta \xi_A = D\varepsilon'_A   + e_2 \Gamma^a \psi_A  \Lambda_a + e_3 \Gamma^{ab}\psi_A \Lambda_{ab} + \nonumber \\
 + e_4 \psi_B \sigma^{x \vert B}_{\;\;\;\;\;A} \Lambda^x + e_5 \Gamma^{abc}\psi_B \sigma^{x \vert B}_{\;\;\;\;\;A} \Lambda^{x\vert}_{\;\;\; abc} + e_6 \Gamma^{a_1...a_5}\psi_A \Lambda_{a_1...a_5}\,,\\
\delta \eta_A = D\varepsilon''_A + l_2 \Gamma^a \psi_A \Lambda_a + l_3 \Gamma^{ab}\psi_A \Lambda_{ab} + \nonumber \\
 +l_4 \psi_B \sigma^{x \vert B}_{\;\;\;\;\;A} \Lambda^x + l_5 \Gamma^{abc}\psi_B \sigma^{x \vert B}_{\;\;\;\;\;A} \Lambda^{x\vert}_{\;\;\; abc} + l_6 \Gamma^{a_1...a_5}\psi_A \Lambda_{a_1...a_5} \,,
\end{array} \right.
 \label{gauge7hf}
\end{eqnarray}
where the parameters  $\Lambda_{i...}$ appearing in (\ref{gauge7hb}) are arbitrary Lorentz (and/or $SU(2)$) valued 0-forms while $\varepsilon'_A, \varepsilon''_A$ in (\ref{gauge7hf}) are arbitrary spinor parameters.

\section{Relation with eleven dimensional Supergravity}\label{11d7d}\label{include3}
The hidden super-Lie algebra discussed in Section \ref{7D} is the most general one for the $D=7$, $\mathcal{N}=2$ supergravity.
Actually, we expect that, for special choices of the parameters, the above structure could be retrieved by dimensional reduction of the eleven dimensional theory, discussed in Section \ref{11D}, in the case where four of the eleven dimensional space-time directions belong to a four-dimensional compact manifold preserving one-half of the supercharges.

The dimensional reduction of eleven dimensional supergravity on an orbifold $T^4/\mathbb{Z}_2$, to the minimal $D=7$ theory, was explicitly performed in \cite{Fre':2015lds}. There, it was pointed out that the  minimal $D=7$ supergravity theory can be obtained as a truncation of the dimensional reduction of $D=11$ supergravity on a torus $T^4 $ (that would gives the maximal $D=7$ theory), where the $SO(4)=SO(3)_+\times SO(3)_-$ holonomy on the internal manifold, is truncated to $SO(3)_+$, so that in the truncation only the reduced fields which are $SO(3)_-$-singlets are retained.

 As far as the fermionic fields are concerned, the truncation selects only 16 out of the 32 components of the eleven dimensional Majorana spinors, described by    pseudo-Majorana spinors  valued in the $SU(2)= SO(3)_+$ seven dimensional  R-symmetry. In particular, the eleven dimensional gravitino 1-form $\Psi$ becomes, in $D=7$:
 \begin{eqnarray}
 \Psi &\to& \psi_A \,,\qquad A=1,2\,.
 \end{eqnarray}
As far as the bosonic fields are concerned, let us  parametrize the Lie algebra of $SO(4)$, the holonomy group of the internal manifold, in terms of the four dimensional `t Hooft matrices $J^{x\,\pm}_{ij}$, where $x=1,2,3$, $i,j,\cdots =1,\cdots ,4$ (their properties  are recalled in Appendix \ref{gammamat}).  The truncation corresponds to drop out the contributions proportional to $J^{x\,-}_{ij}\in SO(3)_-$ in the decomposition of the eleven dimensional bosonic forms to seven dimensions, so that:
\begin{eqnarray}
A^{(3)} &\to& B^{(3)} + A^x \wedge J^{x\,+}_{ij} V^i\wedge V^j\label{117a3}\\
B^{(6)} &\to& B^{(6)} + A^x_{(4)} \wedge J^{x\,+}_{ij} V^i\wedge V^j- 8 B^{(2)}\wedge \Omega^{(4)}\label{117b6}
\end{eqnarray}
where  $V^i$ are the vielbein of the compact manifold and $\Omega^{(4)}=\frac 1{4!}V^{i_1}\wedge\cdots  V^{i_4}\epsilon_{i_1\cdots i_4}$ denotes its volume form.

Next we consider the dimensional reduction  of the Lorentz-valued 1-forms $ \{B_{\hat a\hat b},B_{\hat a_1 \cdots \hat a_5}\}$ of eq. (\ref{sigma11d}), defining the super-Lie algebra hidden in the FDA in $D=11$, to the minimal $D=7$ theory.
We first observe that  comparison of the $D=11$ to the $D=7$ theories would generically require to consider the version of the seven dimensional theory which includes the 1-form $B_{a_1\cdots a_5}$, that in seven dimensions is associated with the (non-dynamical) volume-form $dB^{(6)}$.
Indeed by straightforward  dimensional reduction we obtain:
\begin{eqnarray}
B_{\hat a\hat b} &\to& \left\{
                                \begin{array}{ll}
                                 B_{ab} \\
                                  A^x \,J^{x\,+}_{ij}
                                \end{array}
                              \right. \label{babi}\\
 B_{\hat a_1\cdots \hat a_5} &\to& \left\{
                                \begin{array}{ll}
                                 B_{a_1\cdots a_5} \\
                                 - \frac{3\ii }{2 } A^x_{abc} \,J^{x\,+}_{ij}   \\
                                 - B_a \epsilon_{i_1\cdots i_4}
                                \end{array}
                              \right.\label{bab5}
\end{eqnarray}
where ($\hat a=0,1,\cdots 10$, $a=0,1,\cdots 6$, $i=7,\cdots 10$). Note that to neglect
$B_{a_1\cdots a_5}$ would imply, for consistency of the dimensional reduction, to  drop out also all the other forms in (\ref{bab5}).

As it was observed previously, the  hidden superalgebra in $D=11$  was obtained in \cite{D'Auria:1982nx} by parametrizing only the 3-form $A^{(3)}$ in terms of 1-forms, while the parametrization of the Hodge-dual potential $B^{(6)}$ was not considered there. For this reason we are going to compare  the dimensional reduction of $D=11$ fields considering only the fields appearing in the parametrization of the 3-form.
Considering the fact that the $D=7$ field $B^{(2)}$ descends from the $D=11$ 6-form $B^{(6)}$ (see eq. (\ref{117b6})), comparison of the two theories could shed some light on the parametrization of the $D=11$ form $B^{(6)}$ and  then in the full hidden superalgebra of the $D=11$ theory, since  we cannot get any direct information on the parametrization of  $B^{(6)}$ from the results of \cite{D'Auria:1982nx} reviewed in Section \ref{11D}. In particular, the analysis given in Section \ref{7D} shows that the full hidden super algebra in $D=7$ also includes a second nilpotent spin-3/2 field appearing in the parametrization of $B^{(2)}$, see eq. (\ref{B2par}). As $B^{(2)}$ is a descendent of $B^{(6)}$ from eleven to seven dimensions, this could suggest that considering also the parametrization of $B^{(6)}$ in the analysis of the $D=11$ hidden structure, would amount to include one extra nilpotent fermionic 1-form $\eta'$. A verification of this conjecture by an explicit calculation is left to a future investigation.


Let us quote
 the   set of relations that we found between the $D=7$ and $D=11$ structure constants:
\begin{align} \label{rele117}
& e_1 = \ii E_1, \;\;\; e_2 = -360 E_3, \;\;\; e_3 =E_2, \nonumber \\
& e_4 = 4 \ii E_2, \;\;\; e_5 =120E_3 , \;\;\; e_6 = \ii E_3\\.
\end{align}
The corresponding relation between the coefficients in the parametrizations of the 3-form are:
\begin{align}\label{t117}
& \tau_0 =1, \;\;\; \tau_1=0 , \;\;\; \tau_2 =-\frac{3}{8} T_2, \;\;\; \tau_3=\frac{1}{2}T_1 , \;\;\; \tau_4=7200 T_3 , \;\;\;
\tau_5 =-1296 T_4 , \nonumber \\
& \tau_6= - 216 T_2, \;\;\; \tau_7 =144 T_2 , \;\;\; \tau_8 = -4T_1, \;\;\; \tau_9 =216\times 180 T_4 , \;\;\; \tau_{10}=T_2 , \nonumber \\
& \tau_{11}=0, \;\;\; \tau_{12}=-240 T_3 , \;\;\; \tau_{13}=0, \;\;\; \tau_{14}=1944 T_4 .
\end{align}
In particular, we note that in the dimensional reduced theory $\tau_1=0$, $\tau_{11}=0$, and $\tau_{13}=0$.

Curiously enough, requiring that the set of coefficients (\ref{rele117}) and (\ref{t117}) satisfy the general relations (\ref{taus}) of the seven dimensional theory, implies the condition $T_0=1$ on the set of coefficients of the $D=11$ theory, thus selecting the particular solution (\ref{t01}) originally found in \cite{D'Auria:1982nx}.

We finally write down the  hidden superalgebra in the $D=7$ theory obtained by dimensional reduction from $D=11$:
\begin{align}
& \lbrace Q^A, \overline{Q}_B \rbrace = - \ii \Gamma^a \left(  P_a+ \eta_{ab}Z^b \right) \delta^A_{\; B} - \frac{1}{2} \Gamma_{ab}Z^{ab}\delta^A_{\; B}-\sigma^{x \vert A}_{\;\;\;\;\; B}\left(\ii T^x + \frac{1}{18} \Gamma_{abc}T^{x \vert \; abc}\right), \\
& [Q_A, P_a]= -2 \ii \begin{pmatrix}
5E_2 \\
0
\end{pmatrix} \Gamma_a Q'_A, \\
& [Q_A, Z^a]=-720 \begin{pmatrix}
E_2/48 \\
E_2/72
\end{pmatrix} \Gamma^a  Q'_A, \\
& [Q_A, Z^{ab}]=-4E_2 \Gamma^{ab} Q'_A, \\
& [Q_A, T^x]= - 8 \ii E_2 \sigma^{x \vert B}_{\;\;\;\;\;A} Q'_B , \\
& [Q_A, T^{x \vert abc}]= -1440\begin{pmatrix}
E_2/48 \\
E_2/72
\end{pmatrix}  \Gamma^{abc}\sigma^{x \vert B}_{\;\;\;\;\;A}Q'_B, \\
& [Q_A, Z^{a_1...a_5}] = -2(5!)\ii\begin{pmatrix}
E_2/48 \\
E_2/72
\end{pmatrix}  \Gamma^{a_1...a_5} Q'_A\,.
\end{align}
We see that there are indeed two inequivalent solutions, distinguished by the set of structure constants involving $Q'_A$. In particular the second one features the peculiarity that the commutator $[Q_A, P_a]$ vanishes in corerespondence of the solution $e_1=E_1=0$.
We see that this second solution has a special meaning in the $D=7$ theory: It can be obtained as a special case if we further
require the following identification to hold in the seven dimensional theory:
\begin{equation}
B^{a_1...a_5}= \frac{1}{2} B_{ab}\epsilon^{a_1...a_5ab}\label{kfields}\,.
\end{equation}
The identification is possible in $D=7$ due to the actual degeneration of the Lorentz-index structure for the two 1-forms in (\ref{kfields}). However, in the parent $D=11$ theory the two 1-forms that get identified through (\ref{kfields}) are associated with the mutually non-local exterior forms $A^{(3)}$ and $B^{(6)}$. We speculate that the absence of the coupling of the translation generator to $Q'$ in this case could  possibly be related to the intrinsically topological $D=11$ structure inherent in the association (\ref{kfields}).

\section{Concluding Remarks} \label{concl}
In this paper we have reconsidered the hidden superalgebra structure that underlies  supergravity theories in  space-time dimensions $D>5$ (and in general the supersymmetric theories involving $p$-form gauge fields with $p>1$), first introduced in \cite{D'Auria:1982nx} in the D=11 theory. It generalizes the supersymmetry algebra  to include  the set of almost-central charges (carrying Lorentz indices) which are currently associated with $(p-1)$-brane charges.
We  focussed in particular on the role played by the nilpotent spinor charges naturally appearing in the hidden superalgebra when constructed in the geometrical approach, showing that such extra charges, besides allowing the closure of the algebra, are also  necessary in order for the  FDA to be supersymmetric and gauge invariant on superspace.

Considering in detail the D=11 case, we clarified the physical interpretation of the spinor 1-form field dual to the nilpotent spinor charge: it is not a physical field in superspace, its differential being parametrized  in an enlarged superspace which includes the almost-central charges as bosonic tangent space generators, besides the supervielbein $\{V^a,\psi^\alpha\}$.
Precisely because of this feature, it guarantees that instead the 1-forms dual to the almost central charges are genuine abelian gauge fields whose generators, together with the nilpotent fermionic generators, close an abelian ideal of the supergroup.

As the generators of the hidden super Lie algebra span the tangent space of a supergroup manifold, then in our geometrical approach the fields are naturally defined in an enlarged manifold corresponding to the supergroup manifold, where all the invariances of the FDA are diffeomorphisms, generated by Lie derivatives. The spinor 1-form allows, in a dynamical way, the diffeomorphisms in the directions spanned by the almost central charges to be in particular gauge transformations, so that one obtains the ordinary superspace  as the quotient of the supergroup over the fiber subgroup of gauge transformations.

We have further considered a lower dimensional case,
with the aim to investigate   a possible enlargement of the hidden supergroup structure found in D=11, focussing in particular on the minimal D=7 FDA. Indeed, in that case we were able to parametrize in terms of 1-forms the couple of mutually non-local forms $B^{(2)}$ and $B^{(3)}$. An analogous investigation in D=11 would have required the knowledge of the explicit parametrization of $B^{(6)}$, which is mutually non-local with $A^{(3)}$, but which at the moment has not yet been worked out.
In the seven dimensional case we found that two nilpotent spinor 1-forms are required to find the most general hidden Lie superalgebra. However, as was to be expected, in this case we found  that two subalgebras exist, where only one  spinor, parametrizing only one of the two mutually non-local $p$-forms, is present. We called them Lagrangian subalgebras, since they should correspond to the expected symmetries of a lagrangian description of the theory in terms of 1-forms, or, for the corresponding FDA, to the presence of either $B^{(2)}$ or $B^{(3)}$ in the  lagrangian.

\vskip 5mm
The above results point out to the possible existence of an enlargement also of the D=11  hidden superalgebra, associated with further spinor 1-forms in the parametrization of $B^{(6)}$. This possibility is currently under investigation.

Our results could be extended in several directions which are left to future investigation.

 A relevant issue is the analysis of the hidden structure once  gauge charges are included in the FDA. Moreover, it would be interesting to consider the
dynamical theory based on the 1-form formulation of the supersymmetric FDA, including coupling to matter and more generally a gauging of the theory. A lagrangian description of the interacting theory should be based on one of the Lagrangian subalgebras of the relevant hidden supergroup.
Finally, we observe that the framework worked out in this paper is naturally related to the formulation of double field theory and its generalizations.
As we have seen, the consistency of our framework is implemented dynamically by the very presence of the nilpotent spinor generators in the hidden subalgebra, so that we are led to
 conjecture that the consistency constraints required in double field theory
could be proficiently expressed in our geometrical framework.

\section{Acknowledgements}
We have benefited of stimulating discussions with our friend  Mario Trigiante and we thank him for a critical reading of the manuscript.

\appendix

\section{The explicit solution for $A^{(3)}$ in $D=11$}\label{coeff11D}

In $D=11$, requiring consistency of the parametrization of the $3$-form $A^{(3)}$, see equation (\ref{29}), the following set of equations must be satisfied
\begin{equation} \label{cond11}
\left\{
\begin{array}{l}
T_0-2 S_1 E_1-1=0\cr
T_0-2 S_1 E_2 -2 S_2 E_1=0\cr
3 T_1-8 S_2 E_2=0\cr
T_2+10 S_2 E_3+10 S_3 E_2=0\cr
120 T_3-S_3 E_1-S_1 E_3=0\cr
T_2+1200 S_3 E_3 =0 \cr
T_3-2S_3 E_3=0\cr
9T_4+10 S_3 E_3=0\cr
S_1+10 S_2-720 S_3=0
\end{array}
\right.
\end{equation}
while the integrability condition $D^2 \eta=0$ further implies:
\begin{equation}
E_1+10 E_2-720 E_3 = 0 .
\end{equation}
Here we have also correct some misprints appearing in \cite{D'Auria:1982nx} and \cite{Bandos:2004xw}. This system is solved by the relations (\ref{11dsol}).

In \cite{D'Auria:1982nx} the first coefficient $T_0$ was arbitrarily fixed to $T_0=1$; if we then fix the normalization $T_0=1$ in our system, we get two distinct solutions, depending on the parameter $E_2$ (which just fixes the normalization of $\eta$):
\begin{eqnarray}
T_0 &=& 1 , \;\;\; T_1 \; = \; \frac{4}{15} , \;\;\; T_2 \;=\; -\frac{5}{144}, \;\;\; T_3 \;=\; \frac{1}{17280}, \;\;\; T_4 \;=\; -\frac{1}{31104} ,  \label{t01}\\
S_1 &=& \begin{pmatrix}
0 \\
\frac{1}{2E_2}
\end{pmatrix}  , \;\;\; S_2 \;=\;\frac{1}{10 E_2}, \;\;\; S_3\;=\; \begin{pmatrix}
\frac{1}{720 E_2}  \\
\frac{1}{480 E_2}
\end{pmatrix} , \;\;\; E_1 \;=\; \begin{pmatrix}
5E_2 \\
0
\end{pmatrix} , \;\;\; E_3 \;=\; \begin{pmatrix}
\frac{E_2}{48} \\
\frac{E_2}{72}
\end{pmatrix} . \nonumber
\end{eqnarray}

\section{The explicit solution for $B^{(2)}$ and $B^{(3)}$ in $D=7$}\label{coeff}

As far as the parametrization of  $B^{(2)}$ and $B^{(3)}$ are concerned we distinguish between the case with $B_{a_1\cdots a_5}=0$ and $B_{a_1\cdots a_5}\neq 0$.

\subsection{Coefficients in the case $B_{a_1\cdots a_5}=0$}

\begin{itemize}
\item {\bf Coefficients in the parametrization of $ {B}^{(2)}$}

The coefficients are given by:
\begin{align}
& \sigma = 2 \ii l_2 \tau , \;\;\; l_1 = \frac{\ii}{2\tau}\left(-1+2 \ii l_2 \tau \right), \;\;\;  l_3 = 0, \;\;\; l_4 = \frac{\ii}{2 \tau}, \;\;\; l_5=0, \; \, .
\end{align}

\item {\bf Coefficients in the parametrization of $ {B}^{(3)}$}

If we factorize
\begin{equation}
\frac{e_5}{\sigma_5}={H}
\equiv - 2 \left[{e_1}-2 \ii \left( {e_3}+\frac{\ii}{2} {e_2}\right)\right] \left[{e_1}-2 \ii \left(\frac{\ii}{2} {e_2}+ {e_5}\right)\right],
\end{equation}
we can write the coefficients in the following form:
\begin{align}\label{b3coe}
&  {\tau_0} = 8\left[\frac{\ii}{2} e_1\left(e_3-e_5\right)+\left(\frac{\ii}{2} e_2+ e_3\right)\left(\frac{\ii}{3} e_2+ e_5 \right)\right]\frac{\sigma_5}{e_5}, \nonumber \\
& {\tau_1}= -8 \ii e_2\left(\frac{\ii}{2} e_2+e_5\right) \frac{\sigma_5}{e_5}, \;\;\;  {\tau_2}= -2 e_2^2 \frac{\sigma_5}{e_5}, \;\;\; {\tau_3}= -\frac{16 e_3\left(\frac{1}{2} e_3-e_5 \right)}{3}\frac{\sigma_5}{e_5} , \nonumber \\
& {\tau_4}=2\left(\frac{\ii}{2} e_2+e_5 \right)\sigma_5 , \;\;\;  {\tau_5}= -\ii e_2\sigma_5 , \;\;\; {\tau_6}= 36 e_5\sigma_5 , \nonumber \\
&    {\tau_7}= - 12 e_5 \sigma_5 , \;\;\; {\tau_8}= \frac{2}{3} e_4\left[{e_1}-2 \ii \left(-3
   {e_3}+\frac{\ii}{2} {e_2}+ 3 {e_5} \gamma \right)\right] \frac{\sigma_5}{e_5} , \;\;\;   {\tau_9}= - 3 e_5 \sigma_5 , \nonumber \\
& {\sigma_1}= \left[-e_1+4 \ii \left(\frac{\ii}{2} e_2+e_5 \right)\right] \frac{\sigma_5}{e_5} , \;\;\; {\sigma_2}= e_2 \frac{\sigma_5}{e_5}, \nonumber \\
&   {\sigma_3}= - 2\left(\frac{1}{2} e_3 -e_5\right) \frac{\sigma_5}{e_5}, \;\;\;     {\sigma_4}= - e_4 \frac{\sigma_5}{e_5}, \;\;\;  {\sigma_5}=  \frac{e_5}{H}, \nonumber \\
&   {e_4}= -{e_1}+2 \ii \left(-3 {e_3}+\frac{\ii}{2} {e_2}+ 5 {e_5} \right)\, .
\end{align}
In the relations above, the set of coefficients $\{\sigma_i\}$, that multiply the fermion bilinears in the parametrization of $B^{(3)}$, are given in terms of the structure constants  $\{e_i\}$ appearing in $D\xi_A$. It is noteworthy that the inverse transformation expressing the  $\{e_i\}$  in terms of the $\{\sigma_i\}$ has exactly the same form, since the system of equations is completely symmetric in the interchange of them.
\end{itemize}

\subsection{Coefficients in the case $B_{a_1\cdots a_5}\neq 0$}

\begin{itemize}
\item {\bf Coefficients in the parametrization of  $ {B}^{(2)}$}
\begin{align}
& \sigma = 2 \ii l_2 \tau , \;\;\; l_1 = \frac{\ii}{2\tau}\left(-1+2 \ii l_2 \tau \right), \;\;\;  l_3 = - \frac{60 \chi }{\tau}, \nonumber \\
& l_4 = \frac{\ii}{2 \tau}, \;\;\; l_5 = 0, \;\;\; l_6 =\ii \frac{\chi}{\tau} \, .
\end{align}

\item {\bf Coefficients in the parametrization of $ {B}^{(3)}$}

If we factorize
\begin{equation}
\frac{e_5}{\sigma_5}=\hat{H}
\equiv - 2 \left[{e_1}-2 \ii \left( {e_3}+\frac{\ii}{2} {e_2}-60 \ii e_6\right)\right] \left[{e_1}-2 \ii \left(\frac{\ii}{2} {e_2}+ {e_5} \right)\right],
\end{equation}
we can write the coefficients in the following form:

\begin{align}
& \tau_0 = \frac{2 e_1[e_1-4\ii(\frac{\ii}{2} e_2+e_5 )]}{ \hat{H}} + 120 \tau_{11}+1, \nonumber \\
& \tau_1 =-4\left[\frac{4 (\frac{\ii}{2} {e_2}+ {e_5} ) (\frac{\ii}{2} {e_2}-60\ii {e_6} )}{\hat{H}}+ 60 \tau_{11} + \frac{360 e_1 e_6 ( \frac{1}{6}{e_5}+20\ii {e_6} )}{\hat{H}(\frac{1}{2} {e_3}+90 \ii {e_6} )}\right] , \nonumber \\
& \tau_2 =120\left[ \tau_{11} - \frac{\ii {e_6} (\ii {e_1}+2 \ii
  {e_2}+4 {e_5}) -\ii e_2(\frac{\ii}{2} e_2-30\ii e_6 )}{ \hat{H}}\right]+ \nonumber \\
& \;\;\;\;\;\;\;  +\frac{120  ({e_1}- {e_2})
   {e_6} (-\frac{1}{2} {e_3}+ {e_5} +30\ii {e_6} ) }{\hat{H}(\frac{1}{2} {e_3}+90 \ii {e_6} )} , \nonumber \\
& \tau_3 =   -\frac{16 {e_3} (\frac{1}{2} {e_3}- {e_5} -30\ii {e_6} ) (\frac{1}{2} {e_3}+60\ii {e_6} )}{3  \hat{H} (\frac{1}{2} {e_3}+90\ii {e_6})}, \;\;\; \tau_4 =  \frac{2{e_5} (\frac{\ii}{2} {e_2}+ {e_5} )}{\hat{H}}, \;\;\; \tau_5 =  -\frac{\ii {e_2} {e_5}}{\hat{H}}, \nonumber \\
& \tau_6 =  \frac{216 {e_5} [60\ii {e_6}  (\frac{1}{6}{e_5} +5\ii {e_6})+\frac{1}{2} {e_3} (\frac{1}{6}{e_5} +10\ii {e_6} )]}{\hat{H} (\frac{1}{2} {e_3}+60\ii {e_6})}, \;\;\; \tau_7 =   -\frac{12 e_5^2}{\hat{H}}, \nonumber \\
& \tau_8 =  - \frac{2}{3} \frac{[{e_1}-2 \ii (\frac{\ii}{2} {e_2}-3 {e_3}+3 {e_5} +180\ii {e_6} )][{e_1}-2 \ii (\frac{\ii}{2}
   {e_2}-3 {e_3}+5 {e_5}+180 \ii {e_6} )]}{ \hat{H}}, \nonumber \\
& \tau_9 =   -\frac{3{e_5}^2}{\hat{H}}, \;\;\; \tau_{10} =  -\frac{1200 {e_6}^2 (-\frac{1}{2} {e_3}+ {e_5}   + 30 \ii {e_6} )}{\hat{H} (\frac{1}{2} {e_3}+90\ii {e_6})}, \nonumber \\
& \tau_{11} = \tau_{11}, \;\;\; \tau_{12} =  \frac{4\ii{e_6} [3 \ii {e_1} (\frac{1}{6}{e_5} +20\ii {e_6} )+2 (\frac{\ii}{2} {e_2}+ {e_5} ) (\frac{1}{2} {e_3}+90\ii {e_6} )]}{\hat{H}(\frac{1}{2} {e_3}+90\ii {e_6} )} - 2 \tau_{11}, \nonumber \\
& \tau_{13} = \frac{-2\ii {e_6} [\frac{\ii}{2} ({e_1}- {e_2})
  {e_5} +2 {e_3} (\frac{\ii}{2} {e_2}+\frac{1}{2} {e_5})+60\ii {e_6} (\ii {e_1}+2\ii {e_2}+3 {e_5} ) ]}{\hat{H}(\frac{1}{2} {e_3}+90\ii {e_6} )} +\tau_{11}, \nonumber \\
& \tau_{14} = \frac{18 \ii e_5 e_6(\frac{1}{2} {e_3}-\frac{1}{2} {e_5} +30\ii
  {e_6} )}{\hat{H}(\frac{1}{2} {e_3}+90\ii {e_6} )} , \nonumber \\
& \sigma_1 = -\frac{{e_1}-4 \ii (\frac{\ii}{2}
   {e_2}+ {e_5} )}{\hat{H}} , \;\;\; \sigma_2 = \frac{e_2}{\hat{H}}, \;\;\; \sigma_3 = - \frac{ 2(\frac{1}{2}    {e_3}- {e_5} -30\ii {e_6} ) (\frac{1}{2} {e_3}+60\ii {e_6} )}{\hat{H} (\frac{1}{2} {e_3}+90\ii {e_6} )}, \nonumber \\
& \sigma_4 =   \frac{{e_1}-2 \ii (\frac{\ii}{2} {e_2}-3
   {e_3}+3 {e_5} +180\ii {e_6})}{\hat{H}}, \;\;\; \sigma_5 = \frac{e_5}{\hat{H}}, \;\;\; \sigma_6 = \frac{e_6(\frac{1}{2} {e_3}-{e_5} - 30\ii {e_6} )}{\hat{H}(\frac{1}{2} {e_3}+90\ii {e_6})}, \nonumber \\
&  e_4 = -e_1+2\ii \left(-3 e_3+\frac{\ii}{2} e_2 +5 e_5 + 180\ii e_6 \right) \, .
\end{align}\label{coeff7gen}
\end{itemize}

\section{Dimensional reduction of the gamma matrices}\label{gammamat}

In this section we write the dimensional reduction of the gamma matrices from $D=11$ to $D=7$ dimensions.
We decompose the gamma matrices in eleven dimensions (hatted ones) in the following way:
\begin{equation}
\hat{\Gamma}_{\hat{a}} \rightarrow  \left\{
\begin{aligned}
& 4D \;\;\; \Gamma_i , \\
& 7D \;\;\; \Gamma_a ,
\end{aligned} \right.
\end{equation}
where $\hat{a}=0,...,10$, $a=0,...,6$, and $i=7,8,9,10$.
Then we can write the following decomposition:
\begin{align}
& \Gamma_i = \bfone_{4}\otimes \gamma_i , \\
& \Gamma_a= \Gamma_a \otimes \gamma_5 ,
\end{align}
where
\begin{equation}
\gamma_5 = \begin{pmatrix}
\delta_{A}^{\; B} & 0 \\
0 & -\delta_{A'}^{\; B'}
\end{pmatrix} , \;\;\;\;\; \gamma^5 = \bfone_4,
\end{equation}
and
\begin{equation}
\gamma_i = \begin{pmatrix}
0 & (\gamma_i)_{A}^{\; A'} \\
(\gamma_i)_{A'}^{\; A} & 0
\end{pmatrix} , \;\;\;\;\; \lbrace \gamma_i , \gamma_j \rbrace = 2 \eta_{ij}= - 2 \delta_{ij},
\end{equation}
where $i,j,...$ are the internal index running from $7$ to $10$ and we are using a mostly minus Minkowski metric.
Thus we can write:
\begin{align}
& \Gamma_a = \begin{pmatrix}
(\Gamma_a)_\alpha^{\; \beta}\delta_A^{\; B} & \mathbf{0} \\
\mathbf{0} & -(\Gamma_a)_\alpha^{\; \beta}\delta_{A'}^{\; B'}
\end{pmatrix} , \\
& \Gamma_i = \begin{pmatrix}
\mathbf{0} & (\gamma_i)_A^{\; A'}\delta_\alpha^{\; \beta} \\
(\gamma_i)_{A'}^{\; A}\delta_{\alpha}^{\; \beta} & \mathbf{0}
\end{pmatrix} .
\end{align}

\subsection{Properties of the 't Hooft matrices}

The self-dual and antiself-dual 't Hooft matrices satisfy the quaternionic algebra:
\begin{align}
& J^{\pm \vert x}J^{\pm \vert y} = - \delta^{xy}\bfone_{4\times 4}+ \epsilon^{xyz}J^{\pm \vert z}, \\
& J^{\pm\vert x}_{ab} = \pm \frac{1}{2}\epsilon_{abcd}J^{\pm \vert x}_{cd}, \\
& [J^{+\vert x},J^{-\vert y}]=0, \;\;\; \forall \; x,\;y,
\end{align}
from which it follows
\begin{equation}
\text{Tr}(J^{x}_{rs}J^{y}_{st}J^{z}_{tr})= \text{Tr}(\epsilon^{xyz'}J^{z'}J^{z})= \text{Tr}(-\epsilon^{xyz'}\delta^{z z'}\bfone_4)= -4 \epsilon^{xyz}.
\end{equation}

\section{Fierz identities and irreducible representations}\label{fierz}

\subsection{3-gravitino Irreducible Representations in $D=11$}
The gravitino 1-form $\Psi_\alpha $, $(\alpha =1,\cdots , 32)$,  of eleven dimensional supergravity is a commuting spinor 1-form belonging to the spinor representation of  ${\rm SO(1,10)} \simeq Spin(32)$.
The symmetric product
$(\alpha , \beta , \gamma)\equiv\Psi_{(\alpha} \wedge \Psi_\beta \wedge \Psi_{\gamma )}$, whose dimension is $\mathbf{5984}$, belongs to the three-times symmetric reducible representation  of $Spin(32)$  :
The Fierz identities amount to decompose the given representation $(\alpha , \beta , \gamma)$ into irrepses of $Spin(32)$.
One obtains:
\begin{eqnarray}
\mathbf{5984} \to \mathbf{32}+\mathbf{320}+\mathbf{1408}+\mathbf{4224}
\end{eqnarray}
and the corresponding irreducible spinor representations of the Lorentz group $SO(1,10)$ will be denoted as follows:
\begin{equation}
\Xi^{(32)} \in \mathbf{32} \,,\quad \Xi^{(320)}_a \in \mathbf{320}\,,\quad \Xi^{(1408)}_{a_1a_2}\in \mathbf{1408}\,,\quad \Xi^{(4224)}_{a_1...a_5}\in \mathbf{4224}\,,
\end{equation}
where the indices $a_1\cdots a_n$ are antisymmetrized, and each of them satisfies $\Gamma^a \Xi_{ab_1\cdots b_n}=0$.
One can easily compute the coefficients of the explicit decomposition into the irreducible basis, obtaining: \cite {Castellani:1991et}, \cite{D'Auria:1982nx}:
\begin{eqnarray}
\Psi \wedge \overline{\Psi} \wedge \Gamma_a \Psi & = & \Xi^{(320)} _a+ \frac{1}{11}\Gamma_a \Xi^{(32)},  \\
\Psi \wedge \overline{\Psi} \Gamma_{a_1 a_2}\Psi & = & \Xi^{(1408)}_{a_1a_2}-\frac{2}{9}\Gamma_{[a_2}\Xi^{(320)}_{a_2]}+\frac{1}{11}\Gamma_{a_1 a_2}\Xi^{(32)},  \\
\Psi \wedge \overline{\Psi}\wedge \Gamma_{a_1...a_5}\Psi & = & \Xi^{(4224)}_{a_1...a_5}+2 \Gamma_{[a_1 a_2 a_3}\Xi^{(1408)}_{a_4a_5]}+ \frac{5}{9}\Gamma_{[a_1...a_4}\Xi^{(320)}_{a_5]}-\frac{1}{77}\Gamma_{a_1...a_5}\Xi^{(32)} .
\end{eqnarray}

\subsection{Irreducible representations in $D=7$}
An analogous decomposition in seven dimensions gives:
\begin{eqnarray}
\psi_C \wedge \overline{\psi}^C \wedge \psi_A & = & \Xi _A,  \\
\psi_A \wedge \overline{\psi}^C \wedge \Gamma^{ab}\psi_C & = & \Xi^{ab} _A - \frac{2}{5}\Gamma^{[a}\Xi^{b]}_A + \frac{2}{7}\Gamma^{ab}\Xi_A ,  \\
\psi_A \wedge \overline{\psi}^C \wedge \Gamma^{a}\psi_C & = & \Xi^a_A + \frac{2}{7}\Gamma^a \Xi_A ,  \\
\psi_{(A} \wedge \overline{\psi}_B \wedge \psi_{C)} & = & \Xi_{(ABC)},  \\
\psi_C \wedge \overline{\psi}^C \wedge \Gamma^{abc}\psi_A & = & \frac{3}{2}\Gamma^{[a}\Xi^{bc]}_A + \frac{9}{10}\Gamma^{[ab}\Xi^{c]}_A - \frac{1}{7}\Gamma^{abc}\Xi_A ,  \\
\psi^C \wedge \overline{\psi}^A \wedge \Gamma^{abc}\psi^B & = & \Xi^{(ABC)\vert abc} + \frac{1}{5}\Gamma^{abc}\Xi^{(ABC)} +  \\
& \; & -\frac{2}{3}\epsilon^{C(A}\left(\frac{3}{2}\Gamma^{[a}\Xi^{bc]\vert B)} + \frac{9}{10}\Gamma^{[ab}\Xi^{c]\vert B)} - \frac{1}{7}\Gamma^{abc}\Xi{^\vert B)} \right) ,  \\
\psi^C \wedge \overline{\psi}^A \wedge \psi^B & = & \Xi^{(ABC)}- \frac{2}{3}\epsilon^{C(A}\Xi^{B)}.
\end{eqnarray}


\begin{thebibliography}{99}

\bibitem{string}
As general references on superstring theory, see
M.~B.~Green, J.~H.~Schwarz and E.~Witten,
  ``Superstring Theory. Vol. 1: Introduction,''
   Vol. 2: Loop Amplitudes, Anomalies And Phenomenology,''
  Cambridge, Uk: Univ. Pr. ( 1987) 596 P. ( Cambridge Monographs On Mathematical Physics).\\
  J.~Polchinski,
  ``String theory. Vol. 1: An introduction to the bosonic string'';
  ``String theory. Vol. 2: Superstring theory and beyond''.


\bibitem{Cremmer:1978km}
  E.~Cremmer, B.~Julia and J.~Scherk,
  ``Supergravity Theory in Eleven-Dimensions,''
  Phys.\ Lett.\ B {\bf 76} (1978) 409.
  doi:10.1016/0370-2693(78)90894-8

\bibitem{D'Auria:1982nx}
  R.~D'Auria and P.~Fr\'e,
  ``Geometric Supergravity in d = 11 and Its Hidden Supergroup,''
  Nucl.\ Phys.\ B {\bf 201} (1982) 101
   Erratum: [Nucl.\ Phys.\ B {\bf 206} (1982) 496].
  doi:10.1016/0550-3213(82)90376-5

\bibitem{Haag:1974qh}
  R.~Haag, J.~T.~Lopuszanski and M.~Sohnius,
  ``All Possible Generators of Supersymmetries of the s Matrix,''
  Nucl.\ Phys.\ B {\bf 88} (1975) 257.
  doi:10.1016/0550-3213(75)90279-5

\bibitem{Witten:1978mh}
  E.~Witten and D.~I.~Olive,
  ``Supersymmetry Algebras That Include Topological Charges,''
  Phys.\ Lett.\ B {\bf 78} (1978) 97.
  doi:10.1016/0370-2693(78)90357-X

  \bibitem{vanHolten:1982mx}
  J.~W.~van Holten and A.~Van Proeyen,
  ``$\mathcal{N}=1$ Supersymmetry Algebras in $D=2$, $D=3$, $D=4$ MOD-$8$,''
  J.\ Phys.\ A {\bf 15} (1982) 3763.
  doi:10.1088/0305-4470/15/12/028

\bibitem{Achucarro:1987nc}
  A.~Achucarro, J.~M.~Evans, P.~K.~Townsend and D.~L.~Wiltshire,
  ``Super p-Branes,''
  Phys.\ Lett.\ B {\bf 198} (1987) 441.
  doi:10.1016/0370-2693(87)90896-3

\bibitem{deAzcarraga:1989mza}
  J.~A.~de Azcarraga, J.~P.~Gauntlett, J.~M.~Izquierdo and P.~K.~Townsend,
  ``Topological Extensions of the Supersymmetry Algebra for Extended Objects,''
  Phys.\ Rev.\ Lett.\  {\bf 63} (1989) 2443.
  doi:10.1103/PhysRevLett.63.2443

\bibitem{Abraham:1990nz}
  E.~R.~C.~Abraham and P.~K.~Townsend,
  ``Intersecting extended objects in supersymmetric field theories,''
  Nucl.\ Phys.\ B {\bf 351} (1991) 313.
  doi:10.1016/0550-3213(91)90093-D

\bibitem{Polchinski:1995mt}
  J.~Polchinski,
  ``Dirichlet Branes and Ramond-Ramond charges,''
  Phys.\ Rev.\ Lett.\  {\bf 75} (1995) 4724
  doi:10.1103/PhysRevLett.75.4724
  [hep-th/9510017].

\bibitem{Hull:1994ys}
  C.~M.~Hull and P.~K.~Townsend,
  ``Unity of superstring dualities,''
  Nucl.\ Phys.\ B {\bf 438} (1995) 109
  doi:10.1016/0550-3213(94)00559-W
  [hep-th/9410167].

\bibitem{Townsend:1995gp}
  P.~K.~Townsend,
  ``P-brane democracy,''
  In *Duff, M.J. (ed.): The world in eleven dimensions* 375-389
  [hep-th/9507048].

\bibitem{Bandos:2004xw}
  I.~A.~Bandos, J.~A.~de Azcarraga, J.~M.~Izquierdo, M.~Picon and O.~Varela,
  ``On the underlying gauge group structure of D=11 supergravity,''
  Phys.\ Lett.\ B {\bf 596} (2004) 145
  doi:10.1016/j.physletb.2004.06.079
  [hep-th/0406020]. I.~A.~Bandos, J.~A.~de Azcarraga, M.~Picon and O.~Varela,
  ``On the formulation of D = 11 supergravity and the composite nature of its three-form gauge field,''
  Annals Phys.\  {\bf 317} (2005) 238
  doi:10.1016/j.aop.2004.11.016
  [hep-th/0409100].



\bibitem{Hitchin:2004ut}
  N.~Hitchin,
``Generalized Calabi-Yau manifolds,''
  Quart.\ J.\ Math.\  {\bf 54} (2003) 281
  [math/0209099 [math-dg]].

\bibitem{Gualtieri:2003dx}
  M.~Gualtieri,
 ``Generalized complex geometry,''
  math/0703298 [math.DG].


 \bibitem{Grana:2005jc}
  M.~Grana,
 ``Flux compactifications in string theory: A Comprehensive review,''
  Phys.\ Rept.\  {\bf 423} (2006) 91
  [hep-th/0509003].


 \bibitem{Hull:2005hk}
  C.~M.~Hull and R.~A.~Reid-Edwards,
  ``Flux compactifications of string theory on twisted tori,''
  Fortsch.\ Phys.\  {\bf 57} (2009) 862
  doi:10.1002/prop.200900076
  [hep-th/0503114].

 \bibitem{Dabholkar:2005ve}
  A.~Dabholkar and C.~Hull,
  ``Generalised T-duality and non-geometric backgrounds,''
  JHEP {\bf 0605} (2006) 009
  doi:10.1088/1126-6708/2006/05/009
  [hep-th/0512005].

\bibitem{Grana:2008yw}
  M.~Grana, R.~Minasian, M.~Petrini and D.~Waldram,
  ``T-duality, Generalized Geometry and Non-Geometric Backgrounds,''
  JHEP {\bf 0904} (2009) 075
  doi:10.1088/1126-6708/2009/04/075
  [arXiv:0807.4527 [hep-th]].



\bibitem{Hull:2007zu}
  C.~M.~Hull,
  ``Generalised Geometry for M-Theory,''
  JHEP {\bf 0707} (2007) 079
  doi:10.1088/1126-6708/2007/07/079
  [hep-th/0701203].



  \bibitem{Pacheco:2008ps}
  P.~P.~Pacheco and D.~Waldram,
 ``M-theory, exceptional generalised geometry and superpotentials,''
  JHEP {\bf 0809} (2008) 123
  doi:10.1088/1126-6708/2008/09/123
  [arXiv:0804.1362 [hep-th]].

\bibitem{Coimbra:2012af}
  A.~Coimbra, C.~Strickland-Constable and D.~Waldram,
  ``Supergravity as Generalised Geometry II: $E_{d(d)} \times \mathbb{R}^+$ and M theory,''
  JHEP {\bf 1403} (2014) 019
  doi:10.1007/JHEP03(2014)019
  [arXiv:1212.1586 [hep-th], arXiv:1212.1586].


\bibitem{Hull:2009mi}
  C.~Hull and B.~Zwiebach,
  ``Double Field Theory,''
  JHEP {\bf 0909} (2009) 099
  doi:10.1088/1126-6708/2009/09/099
  [arXiv:0904.4664 [hep-th]].
  \\ For a recent review, see for example: O.~Hohm, D.~Lüst and B.~Zwiebach,
  ``The Spacetime of Double Field Theory: Review, Remarks, and Outlook,''
  Fortsch.\ Phys.\  {\bf 61} (2013) 926
  doi:10.1002/prop.201300024
  [arXiv:1309.2977 [hep-th]].


\bibitem{Hull:2009zb}
  C.~Hull and B.~Zwiebach,
  ``The Gauge algebra of double field theory and Courant brackets,''
  JHEP {\bf 0909} (2009) 090
  [arXiv:0908.1792 [hep-th]].

  \bibitem{Hohm:2010jy}
  O.~Hohm, C.~Hull and B.~Zwiebach,
  ``Background independent action for double field theory,''
  JHEP {\bf 1007} (2010) 016
  [arXiv:1003.5027 [hep-th]].

  \bibitem{Hohm:2010pp}
  O.~Hohm, C.~Hull and B.~Zwiebach,
 ``Generalized metric formulation of double field theory,''
  JHEP {\bf 1008} (2010) 008
  [arXiv:1006.4823 [hep-th]].

  \bibitem{Hull:2014mxa}
  C.~M.~Hull,
 ``Finite Gauge Transformations and Geometry in Double Field Theory,''
  JHEP {\bf 1504} (2015) 109
  [arXiv:1406.7794 [hep-th]].


\bibitem{Hohm:2013pua}
  O.~Hohm and H.~Samtleben,
  ``Exceptional Form of D=11 Supergravity,''
  Phys.\ Rev.\ Lett.\  {\bf 111} (2013) 231601
  doi:10.1103/PhysRevLett.111.231601
  [arXiv:1308.1673 [hep-th]].

\bibitem{Hohm:2013uia}
  O.~Hohm and H.~Samtleben,
``Exceptional field theory. II. E$_{7(7)}$,''
  Phys.\ Rev.\ D {\bf 89} (2014) 066017
  doi:10.1103/PhysRevD.89.066017
  [arXiv:1312.4542 [hep-th]].

  \bibitem{Hohm:2014qga}
  O.~Hohm and H.~Samtleben,
``Consistent Kaluza-Klein Truncations via Exceptional Field Theory,''
  JHEP {\bf 1501} (2015) 131
  doi:10.1007/JHEP01(2015)131
  [arXiv:1410.8145 [hep-th]].




\bibitem{Sullivan}
D.~Sullivan,
 ``Infinitesimal computations in topology". Publications Math\'ematiques de l'IHES, 47 (1977), p. 269-331

 \bibitem{Castellani:1991et}
  L.~Castellani, R.~D'Auria and P.~Fr\'e,
  ``Supergravity and superstrings: A Geometric perspective,'' Vol. 1 and 2.
 Singapore: World Scientific (1991)

 \bibitem{Townsend:1983kk}
  P.~K.~Townsend and P.~van Nieuwenhuizen,
  ``Gauged Seven-dimensional Supergravity,''
  Phys.\ Lett.\ B {\bf 125} (1983) 41.
  doi:10.1016/0370-2693(83)91230-3

  \bibitem{Salam:1983fa}
  A.~Salam and E.~Sezgin,
  ``SO(4) Gauging of $\mathcal{N}=2$ Supergravity in Seven-dimensions,''
  Phys.\ Lett.\ B {\bf 126} (1983) 295.
  doi:10.1016/0370-2693(83)90167-3

  \bibitem{Bergshoeff:1985mr}
  E.~Bergshoeff, I.~G.~Koh and E.~Sezgin,
  ``{Yang-Mills} / Einstein Supergravity in Seven-dimensions,''
  Phys.\ Rev.\ D {\bf 32} (1985) 1353.
  doi:10.1103/PhysRevD.32.1353

  \bibitem{Fre':2015lds}
   P.~Fr\'e, P.~A.~Grassi, L.~Ravera and M.~Trigiante,
  ``Minimal $D=7$ Supergravity and the supersymmetry of Arnold-Beltrami Flux branes,''
  JHEP {\bf 1606} (2016) 018
  doi:10.1007/JHEP06(2016)018
  [arXiv:1511.06245 [hep-th]].
















\end{thebibliography}
\end{document}